\documentclass[superscriptaddress,pra,twocolumn,showpacs]{revtex4} 

\usepackage{graphicx}
\usepackage{float}
\usepackage{epstopdf}
\usepackage{color}
\usepackage{amsmath}
\usepackage{amssymb}

\begin{document}

\title{Elastic and  inelastic diffraction of fast atoms,\linebreak Debye-Waller factor and M\"{o}ssbauer-Lamb-Dicke regime}
\author{Philippe Roncin and Maxime Debiossac}

\affiliation{Institut des Sciences Mol\'{e}culaires d'Orsay (ISMO), 
	CNRS, Univ. Paris-Sud, Universt\'{e} Paris Saclay, F-91405 Orsay, France}
\date{\today}

\pacs{34.20.Cf,34.35.+a,34.50.-s,34.50.Bw,34.50.Cx,68.49.Bc,79.20.Rf,79.60.Bm}

\begin{abstract} 
	The diffraction of fast atoms at crystal surfaces is ideal for a detailed investigation of the surface electronic density. However, instead of sharp diffraction spots, most experiments show elongated streaks characteristic of inelastic diffraction. This paper describes these inelastic profiles in terms of individual inelastic collisions with surface atoms taking place along the projectile trajectory and leading to vibrational excitation of the local Debye oscillator. A quasi-elastic regime where only one inelastic event contributes is identified as well as a mixed quantum-classical regime were several inelastic collision are involved. These regimes describe a smooth evolution of the scattering profiles from sharp spots to elongated streaks merging progressively into the classical diffusion regime. 
\end{abstract}

\maketitle
\section{Introduction}

The interaction of keV atoms with surfaces has a long history, motivated, in part by industrial applications such as plasma facing materials in tokamaks, but also by the specific behavior of ions and atoms to probe surfaces compared with photons or electrons. 
The combination of grazing incidences and single crystal surfaces has offered perfect conditions to understand a variety of basic physical processes, for instance the resonant coherent excitation\cite{Auth_97} of H atoms by the surface electric field or, reversely the excitation of surface optical phonons by the electric field of the moving projectile ions\cite{Winter_rep_2002,Villette_2000}. 
The interactions of keV ions or atoms with the surface consist, in grazing incidence, of multiple collisions that are well controlled so that individual surface electronic excitations such as excitons\cite{Roncin_98} or trions\cite{Roncin2002} have been identified together with their role in electronic emission from ionic insulator\cite{Roncin_98,Winter_rep_2002}. 
Progressively weaker and weaker interactions were probed resulting in the observation of a quantum behavior illustrated by diffraction features in the scattering profile\cite{Khemliche_patent,Schuller_2007,Rousseau_2007}. 

Grazing incidence fast atom diffraction (GIFAD or FAD) is an extreme surface sensitive technique (see\cite{Winter_PSS_2011} for a review) perfectly suited to probe, in real time and at high temperatures, the surface specific structures known as surface reconstructions\cite{Atkinson_2014}.

Despite elastic diffraction of keV atoms being predicted theoretically \cite{Manson2008,Andreev}, initial experimental diffraction patterns\cite{Khemliche_patent,Schuller_2007,Rousseau_2007} did not consist of sharp diffraction spots located on the Laue circle i.e. corresponding to energy conservation, which is the signature of elastic diffraction.
Later on, several experiments using surfaces with large enough coherence length revealed clear evidence of elastic diffraction\cite{Busch_2012,DebiossacPRB_2014,DebiossacPRL,Lalmi}.

    \begin{figure}	\includegraphics[width=0.9\linewidth]{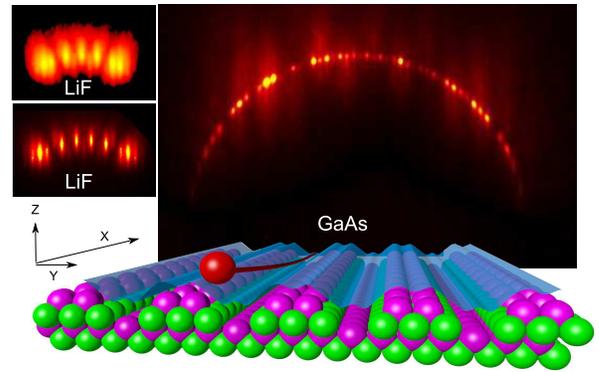}
	\caption{Schematic view of a GIFAD setup. A beam of keV helium atoms interacts at grazing incidence with crystal surface. Here a pristine GaAs surface at $\sim900K$ inside a MBE chamber\cite{DebiossacPRB_2014,Atkinson_2014}. The diffraction pattern is recorded $\sim1$m downstream on a detector. The bright spots sitting on the Laue circle correspond to the elastically scattered intensity. The insets on the left show patterns\cite{Rousseau_2007,Momeni} with larger streaks corresponding to inelastic regimes investigated here.
	\label{fig_Setup}} \end{figure}

This is illustrated in Fig.\ref{fig_Setup} where a typical experimental setup is sketched. A keV ion beam is first neutralized and sent inside a UHV chamber to interact with the surface at angles close to one degree. If the crystal surface is aligned along a low index direction, the detector placed downstream shows a diffraction pattern. The one displayed in Fig.\ref{fig_Setup} corresponds to the $\beta_2 (2\times 4)$ reconstruction of GaAs(001) along the [1-10] direction\cite{DebiossacPRL}. It was recorded inside a molecular beam epitaxy (MBE) vessel using a GaAs surface at high temperature ($\sim580^\circ C$). Also, it has been shown that the quality of the terminal layer is a prerequisite to monitor layer by layer growth dynamics\cite{Atkinson_2014}. 

In most cases the Laue circle clearly visible in Fig.\ref{fig_Setup} for the GaAs surface is not present indicating the lack of energy conservation. This raises two important questions: is the diffraction information impaired in this inelastic regime ? is there something to be learned from these inelastic profiles?

Before addressing these issues, the inelastic regime must be understood better and this paper proposes an approach based on a sudden approximation where individual phonon modes are not included explicitly.

The paper is organized as follows : theoretical models are briefly presented together with well established theory of decoherence using the Debye Waller factor ($DWF$) in spatial and momentum domains. Then the specific conditions of grazing incidence scattering are examined from both spatial and momentum points of view. 
A planar description of the interaction of the atom with the surface is proposed. This leads to a new presentation of the $DWF$ where the classical projectile energy loss determines the elastic scattering probability and suggests  the existence of a new mixed quantum-classical regime. 
A unified description of these different regimes is proposed showing a smooth continuity through the crossover between quantum and classical diffusion. 
From the statistical properties of the individual collisions, the model predicts the angular and energy loss distributions and the associated line shapes of the inelastic diffraction peaks. 
These predictions are then confronted with existing experimental results in the literature. 

\section{Theoretical descriptions}

All theoretical diffraction models for fast atom diffraction start from a rigid surface lattice with atoms standing still at their equilibrium positions. The potential energy landscape is determined by quantum chemistry techniques, density functionals, or model binary potentials. The specificity of grazing angles is accounted for by averaging the actual $3D$ potential energy surface  $V_{3D}(x,y,z)$ along the direction of the fast movement (here $x$ as in Fig.\ref{fig_Setup}) producing a 2D landscape $V_{2D}(y,z)$ where a particle with energy $E_\perp$ diffracts just as in standard thermal energy atom scattering\cite{Rousseau_2007,FariasCPL2004}. The range of validity of this axial channeling approximation has been investigated in detail\cite{Zugarramurdi_2012,Diaz_2016b,danailov}. 
With this energy landscape, several techniques have successfully described the diffraction patterns such as exact wave packet\cite{Rousseau_2007,Aigner_2008,DebiossacPRB_2014}, close coupling\cite{DebiossacPRL} or multi configuration time-dependent Hartree\cite{Diaz_2016b}. 
Other approaches based on Bohmian trajectories\cite{Sanz} or even classical trajectories and semi classical approaches\cite{FariasPRL2004,Gravielle_2014} including specific correction of the rainbow divergence\cite{Gravielle_2014} have shown good agreement with experimental results. 

The simplest model one may think of is the hard corrugated wall approach (HCW). In this model the averaged $2D$ potential energy landscape $V_{2D}(y,z)$ is replaced by a 1D corrugation function $Z_c(y)$ defined by energy conservation $V_{2D}(y,Z_c(y))=E_\perp$ \cite{Armand_1979}. Considering this 1D corrugation function $Z_c(y)$ as a mirror like grating, an optical model is enough to predict the diffracted intensities as a Fourier like transform of  $Z_c(y)$. Elastic diffraction implies that no energy is exchanged with the surface, hence the profile of the diffracted beams are supposed to be the same as the primary beam, in contrast with most experimental observations.

In the case of inelastic diffraction although many experimental results have been demonstrated, no well-established theory is available to analyze them.
 So far, experiments have been interpreted using elastic theories. There has been an attempt to describe observed diffraction results using an elastic wave packet calculation perturbed by random kicks to the wave function\cite{Aigner_2008}. It showed good agreement to inelastic data but the angular profile was not predicted. Instead the angular profile was adjusted by tuning properties of the initial wave packet. Furthermore, this calculation did not account for elastic diffraction and no indication was given how to link both processes. 
Soon after, a general framework based on the transition matrix formalism was proposed in Ref.\cite{Manson2008} to describe both elastic and inelastic processes. This model includes all phonon modes, however it does not provide an easy way to calculate the relevant transition matrix elements. 
The model developed hereafter can be seen as a simplification where the scattering process is expanded in terms of individual elastic or inelastic collisions rather than in terms of individual phonons. Before describing this new model we briefly recall decoherence theory using the Debye-Waller factor.

\section{Coherence and diffraction}

The Debye-Waller factor can be seen as the ratio of the coherent scattered intensity ($I_c$) over the total scattered intensity ($I_0$) of the primary beam
\begin{equation} DWF=\frac{I_c}{I_0}=e^{-q^2\langle z^2 \rangle }  \label{DWF_std} \end{equation}
As usual in quantum mechanics, several interpretations are possible from the standpoint of either real or momentum space.  
 
\subsection{Spatial approach}

\begin{figure}	\includegraphics[width=0.8\linewidth]{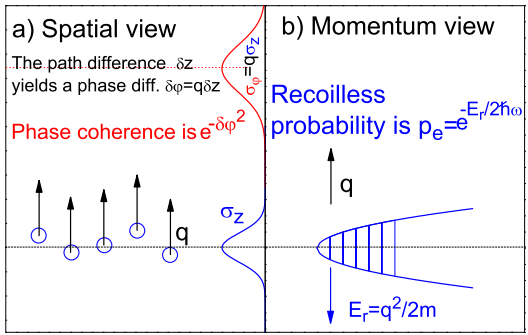}
	\caption{Schematic view of the two approaches to the decoherence due to thermal vibrations. a) the coherence of an ensemble is limited by the spatial spread of the emitters. b) the probability $p_e=$ of recoilless emission from an harmonic oscillator in its ground state $\psi\rangle$ is  $|\langle\psi|  e^{iq z} |\psi\rangle|^2$. Both approaches give identical result.
		\label{fig_DWF}}
    \end{figure}

The compact form of the DWF given in Eq.\ref{DWF_std}, where $q$ is the momentum transfer and $\langle z^2 \rangle$ denotes the thermal mean square displacement of the surface atoms, has a simple geometric interpretation. 
It is related to the path difference between different trajectories leading to the same final scattering angle. 
Bragg conditions correspond to certain directions in space ($\vec{q}$) where incident particles with a wave-vector $\vec{q}$ are scattered off by a periodic array of atoms located at their equilibrium position. A displacement $\delta \vec{r}$ will then give rise to a path difference $\delta \vec{r}$ and a phase shift $\delta \varphi= \vec{q}.\delta\vec{r}$ (see Fig.\ref{fig_DWF}a). Switching to one dimension $z$ for simplicity, a Gaussian distribution of surface atoms characterized by a standard deviation $\sigma_z$ produces a Gaussian phase distribution with standard deviation $\sigma_\varphi=q\sigma_z$. The global coherence of these waves (amplitudes) is $e^{-q^2 \sigma_z^2}$ which is exactly the $DWF$ if one identifies $\langle z^2 \rangle$ to $\sigma_z^2$.

The evaluation of $\langle z^2 \rangle$ is usually performed in the harmonic approximation defined by the frequency $\omega$. For the ground state $\langle   z^2  \rangle  = \frac{\hbar}{2m\omega} $ so that $DWF=exp(-\frac{q^2\hbar}{2m\omega})$. 

\subsection{Momentum approach}

The $DWF$ can also be written in a form where the recoil energy  $E_r=\frac{\hbar^2 q^2}{2m}$ is explicit 
\begin{equation} DWF = e^{\frac{-E_r}{\hbar\omega}}    \label{DWF_Er}.\end{equation}

This form of is better known in the atomic and nuclear physics community as the M\"{o}ssbauer-Lamb-Dicke factor. It gives the fraction of recoilless emission from independent trapped particles. At a single particle level, it is more convenient to define M\"{o}ssbauer-Lamb-Dicke factor as a probability for recoilless emission.
This is illustrated in M\"{o}ssbauer spectroscopy where $\gamma$ rays are absorbed by iron nuclei in a crystal\cite{Nagy}, in spectroscopic line narrowing at high pressure\cite{Dicke_1953} or in Doppler free interaction of light with cold atoms in an optical lattice\cite{Wineland_1987}.

In these cases, there is no interference between emitted waves and Eq.\ref{DWF_Er} can be interpreted as the probability $p_e=|\langle \varPsi| e^{iq z} |\varPsi\rangle|^2$ of exchanging a momentum $q$ with a particle, without changing its energy. Modeling the trap by an harmonic oscillator with resonant frequency $\omega$, the solution is straightforward using the Bloch theorem\cite{Cohen_Tannoudji} 
\begin{equation} \langle  e^{iqz}  \rangle = e^{-\frac{1}{2} q^2 \langle   z^2  \rangle  }=   e^{\frac{-E_r}{2\hbar\omega}}   \label{DWF_overlap}.\end{equation}
With a squared value $|\langle  e^{iqz}  \rangle|^2=e^{\frac{-E_r}{\hbar\omega}}$ identical to the $DWF$ factor.

The recoil energy $E_r$ reported above is the classical kinetic energy lost by the projectile and transferred to the surface atom associated with the momentum $q$ exchanged with the surface atom. 
However, when this $DWF$ probability is close to unity, i.e. in the recoilless Lamb-Dicke regime, the trapped atom does not change momentum. 
The basic laws of physics are preserved since the system is not isolated; the whole crystal or the experimental setup, responsible for the trapping potential, collects the exchanged momentum without recoil energy.

In thermal energy atom scattering (TEAS) both interpretations of the same formula can be given. On one hand elastic diffraction implies that no energy is exchanged with the surface i.e. recoilless reflection and, on the other hand bright diffraction peaks can be observed only if the thermal fluctuations of the scatterer do not destroy the coherence. A specificity of neutron or helium diffraction is that the mass $m_p$ of the projectile is comparable to the mass $m$ of the surface atoms. This means that the Lamb-Dicke regime, or high coherence diffraction, can exist only for projectile kinetic energies on the order of $\hbar\omega$ the surface atoms vibration energy. This explains why diffraction of keV atoms came as a relative surprise even when considering the relative decoupling of motion $\parallel$ and $\perp$ to the surface.

\section{grazing angle, coherence and multiple collisions}
\subsection{ Momentum approach}

\begin{figure}	\includegraphics[width=0.95\linewidth]{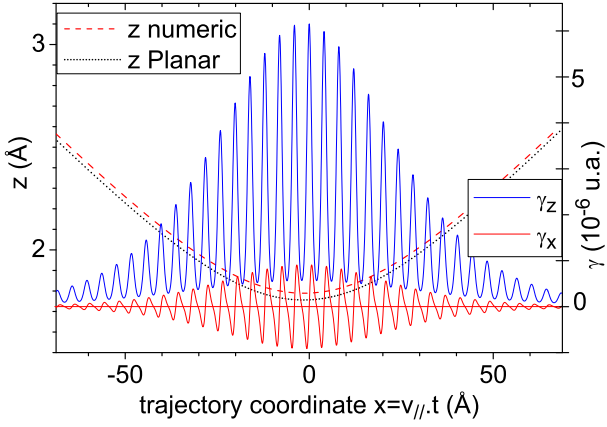}
	\caption{(Color online) Classical trajectory $z(x)$ of a 1keV helium projectile calculated on top of a row of fluorine atoms. Note that the $z$ scale (left) is $\sim$ 100 times the $x$ scale. The smooth trajectory is made of successive localized interactions with the surface atoms as illustrated by the components $\gamma_x, \gamma_z$ of the acceleration along the trajectory (right scale).
		\label{fig_gamma}}
\end{figure}

Taking a rigid LiF lattice and the binary interaction potential published in\cite{DebiossacPRL}, the trajectory of a 1 keV  helium atom impinging at one degree incidence can be integrated numerically. Figure \ref{fig_gamma} shows such a trajectory together with the acceleration $\gamma_x$ along the beam direction and $\gamma_z$ perpendicular to the surface. A peak in the acceleration along z is present each time the projectile flies over a surface atom. $\gamma_x$ oscillates around zero indicating that the slowing down in front of an atom is immediately followed by an acceleration behind, limiting the momentum transfer along $x$. Overall, the integral of $\gamma_x$ tends to zero as noted in\cite{FariasCPL2004,FariasPRL2004,Rousseau_2007,Zugarramurdi_2012} and calculated analytically in\cite{Henkel1994,Debiossac_PRA}. This justifies the use of the axial surface channeling approximation where, schematically the surface egg-carton-like 3D surface corrugation is replaced by a 2D washboard-like surface potential profile. This cancellation of the integral momentum transfer along $x$ does not apply for $\gamma_z$ because all peaks are positive (directed towards the vacuum) and progressively repel the projectile always in the same direction allowing specular reflection. 

\begin{figure}	\includegraphics[width=0.95\linewidth]{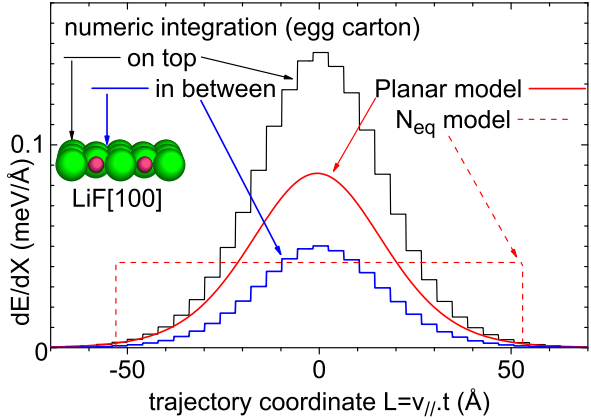} 
	\caption{(Color online) The energy transferred to the surface atoms is estimated by taking the integral under each peak in the acceleration curve $\gamma_z$ of Fig.\ref{fig_gamma} for trajectories on top of the atomic rows (black histogram) or in between two rows (blue). The predictions of the structureless planar model is the quasi gaussian red curve ($\sigma \sim 1.04/\Gamma a$). The $N_{eq}$  model (dashed red ) assumes $N_{eq}$ equivalent lattice sites.
		\label{fig_dEdx}}
\end{figure}

For each binary collision, the momentum transfer can be converted into a virtual recoil energy and are plotted in Fig. \ref{fig_dEdx}. 
The probability $P_e$ that all binary collision are elastic is given by the product of each individual probability $p_e$ ; $P_e=\Pi _{i=1}^{i=N} p_e$. 
Taking the form of Eq.\ref{DWF_Er} for each of these collisions with individual recoil energies $E_{ri}$ for $p_e$, one obtains a form where factorization leads to the sum of the recoil energies $E_{loss}=\Sigma _{i=1}^{i=N}E_{ri}$.

\begin{equation}
P_e=\Pi _{i=1}^{i=N} exp(\frac{-E_{ri}}{\hbar\omega}   )  = exp(\frac{-\Sigma _{i=1}^{i=N}E_{ri}}{\hbar\omega} ) \nonumber 
\end{equation} 

This is similar to  Eq.\ref{DWF_Er} except that here the energy loss results from the sum of each individual energy loss with the surface atoms. 
\begin{equation}
\fbox {$P_e= exp(-\frac{E_{loss}}{\hbar\omega}  )$}        \label{Pe_LD}
\end{equation}

Such a compact form, similar to that of the Lamb-Dicke regime, is new in the grazing incidence context where $E_{loss}$ is the sum of virtual energy transfer over a large number of collisions.

\subsection{Trajectory length, continuous model, and projectile energy loss}

The trajectory reported in Fig.\ref{fig_gamma} is derived from a straightforward integration of the Newtonian equations of motion. 
Each peak in the acceleration curve can be integrated and associated to a given momentum exchange and recoil energy transferred to the surface atoms at each lattice site producing an energy deposition curve. 
Two examples corresponding to 'on-top' and 'in the valley' trajectories are plotted in Fig.\ref{fig_dEdx}. Both display a strong gaussian character with significantly different amplitudes but with a common well defined width. Note that for the grazing angle considered here all trajectories end up on the Laue circle confirming the axial channeling model in the classical regime.

A simpler model can be proposed where the contributions of the binary potentials $V(r)$ are averaged to produce a planar potential $V_p(z)$. Using Moliere-type radial potential $V(r)=\frac{V}{r} e^{-\Gamma r}$ the planar form is $V_p(z)=\frac{2\pi}{\Gamma} n_s V e^{-\Gamma z}$ where  $n_s =1/a^2$  is the surface density with one Fluorine atom per lattice site ($a=2.85\AA = 4.03/\sqrt(2)$)\cite{Debiossac_PRB_2016}.
In this translation invariant potential the movements parallel and perpendicular to the surface are decoupled.  Let us call $v_\parallel = v_{ix}$ and $v_\perp=-v_{iz}$ the initial velocity components parallel and perpendicular to the surface, $\theta = \tan (v_\perp/v_\parallel) \sim v_\perp/v_{\parallel}$, and $z_0$ the turning point such that $V_p(z_0)=E_\perp=E\sin^2\theta$.
The characteristic time $\tau$ for a half turn on the surface depends on the range $1/\Gamma$ of the potential, $\tau\sim 1/\Gamma v_\perp$ so that the interaction length is $L\sim v_\parallel\tau \propto 1/\Gamma \theta$ independent of the projectile mass or energy. 

More precisely, the trajectory $z(t)$ can be integrated analytically as well as its derivative $\dot{z}$ and second derivative $\ddot{z}$ giving the angle $\theta(t) = \dot{z}/ v_\parallel$ and its square $\mu\ddot{z}^2/2 $ ($\mu=m_{proj}/m_{target}$ is the mass ratio) corresponding to an energy deposition curve;

 \begin{equation}\label{dEt} 
	\begin{aligned}
		&z(t)=z_0 + v_\perp t +  \frac{2}{\Gamma} \ln (\frac{1+\exp^{(-\Gamma v_\perp t)} }{2})\\
		&dE(t)=\mu E a \frac{1}{v_\parallel^3}~ \frac{\Gamma^2 v_\perp^4}{4  \cosh^4(\Gamma v_\perp t/2) }\\
		&dE(x)\sim ~ \frac{\mu E a \Gamma^2 \theta^4}{4  \cosh^4(\Gamma \theta x/2) }\\
	\end{aligned}
\end{equation}

The trajectory and energy deposition curves are plotted in Fig.\ref{fig_gamma} and \ref{fig_dEdx} respectively. Compared with their numerical counterpart calculated on top of the fluorine rows or in between, the planar formula shows a comparable width and a magnitude somewhere between 'on-top' and 'in-between'. Eq.\ref{dEt} can be integrated to produce the total energy loss $E_{loss}$ specific to grazing incidence\cite{Manson2008}.

\begin{equation}
\fbox {$E_{loss} = \frac{2}{3} \mu E \Gamma a \theta_{in}^3$}   \label{Eloss}        
\end{equation}

It is interesting to outline the surface effect by comparing the energy loss of Eq.\ref{Eloss} to the energy loss $\delta E_{single}$ expected if only one atom would produce the total deflection $2\theta$=($\theta_{in}+\theta_{out}$);
$\delta E_{single} =\mu E (2\theta)^2$ (small angle formula derived from energy momentum conservation).  
The ratio of these two values indicates\cite{Manson2008} that $E_{loss}$ is $N_{eq}$ times smaller than $\delta E_{single}$ with $N_{eq}$ given by 
\begin{equation}
	N_{eq}= \frac{6}{\Gamma a \theta_{in}}    \label{Neq}        
\end{equation}
Another important parameter is the peak value $\delta E_{max}$ of Eq.\ref{dEt} per lattice unit, corresponding to the central and most violent collision encountered along the trajectory.

\begin{equation}
	\fbox {$\delta E_{max} =~\mu E~\Gamma^2 a^2 \theta_{in}^4 /4$}   \label{DE_Max}        
\end{equation}

\subsection{The equivalent scatterers model}

In the equivalent scatterers model, used hereafter only for illustration purposes, a further simplification is made by considering that $N_{eq}$ successive collisions participate equally, by $\delta\theta_{eq}=2\theta/N_{eq}$ to the total deflection, and to the total energy loss by $E_r=E_{loss}/N_{eq}$.
The contributions of an individual scattering are
\begin{equation}
\delta\theta_{eq}=~\Gamma a \theta^2 /3~,~~E_r=\mu E \Gamma^2 a^2 \theta^4 /9   \label{Eri}        
\end{equation}

Note that with $\theta\sim 1 deg~\sim 1/57$ rad., $\theta^4$ is on the order of $10^{-8}$ underlining that there should always be an angle for which the Lamb-Dicke regime will be reached, \textit{i.e.} where the individual recoil energy $E_r$ is much less than the vibration energy quantum $\hbar\omega$. Of course, this holds only if the surface quality allows such grazing trajectories to develop without encountering topological defects. 

This equivalent scatterers model considers a straight line trajectory of length $L =N_{eq}\times a$ parallel to the surface. 
It is a discrete version of the effective length model used, for instance, to link the observed variation of the neutralization fraction with the angle of incidence to an electron density dependent Auger rates\cite{Rousseau_NIM_2007}. 
Fig.\ref{fig_dEdx} shows that the length defined here is close to twice the fwhm of the energy deposition curve and, consistently, that the effective recoil energy is close to half of the maximum.


\subsection{ Spatial approach}

As recalled in Eq. \ref{DWF_std}, the $DWF$ can be expressed as a function of the spatial fluctuation of the scatterers and interpreted as a dephasing of the scattered waves. At grazing incidence the reflection of the projectile occurs on the rows of well aligned atoms. Considering that each row consists of $N_{eq}$ scatterers\cite{Rousseau_2008,Manson2008}, the thermal position fluctuations of the portion of the probed row, should be reduced to $\sigma_z/\sqrt{N_{eq}}$ where $\sigma_z$ is the position fluctuation of a single surface atom. 

 \begin{equation}
 DWF=\frac{I_c}{I_0}=exp(\frac{-q^2\langle z^2 \rangle}{N_{eq}} )  \label{DWF_Spat}
 \end{equation}

This is identical to Eq.\ref{Eloss} and \ref{Pe_LD} so that the decoherence in GIFAD can also be presented both in spatial and energetic terms. In the elastic diffraction calculation these rows are considered infinite but it is precisely the finite length that allows a simple estimate of the elastic scattering probability via the phase coherence in the $DWF$ adapted to grazing incidence.

\subsection{Temperature and Debye Model}
The simple formulae reported above are valid for isolated ground state harmonic oscillators. They have to be adapted to solids where all the local oscillators are connected together giving rise, in the Debye model, to an increase of $\langle z^2 \rangle$ by a factor $3$ when summing over all phonon contributions\cite{Desjonqueres_94}. The temperature effects are easily accounted for by multiplying the ground state extension $\langle z^2 \rangle$ by $\coth(\frac{T_D}{2T})$ where $T_D$ is the Debye temperature describing the local oscillator; $k_B T_D=\hbar\omega$.
This exact formula, derived from Boltzmann weighting of the harmonic oscillator wave-functions, starts at unity for $T=0$, increases slowly above two for $T=T_D$ and reaches the classical Dulong and Petit limit with a linear behavior above $T_D$. Overall the crude estimate of $\langle   z^2  \rangle$ from an isolated oscillator in Eq.\ref{DWF_overlap} has to be multiplied by $3\coth(\frac{T_D}{2T})$.

\begin{equation} \langle   z^2  \rangle  = \frac{3\hbar}{2m\omega}\coth(\frac{T_D}{2T}) =  \frac{3\hbar^2}{2mk_B T_D}\coth(\frac{T_D}{2T}) \label{z2T} \end{equation}

For an individual event associated with an energy $\delta E$ this gives an elastic probability $p_e$
\begin{equation} p_e= exp(-3\frac{\delta E}{k T_D} \coth(\frac{T_D}{2T}) )  \label{pe} \end{equation}

 and for the entire trajectory
\begin{equation} \fbox {$DWF=P_e= exp(\frac{- 2\mu E ~\Gamma a \theta_{in}^3}{k T_D} \coth(\frac{T_D}{2T})  )$} \label{Pes} \end{equation}

Consistently, using Eq.\ref{pe} in the $N_{eq}$ model gives a constant individual probability

 $p_{eq}= exp(-\frac{ \mu E ~\Gamma^2 a^2 \theta_{in}^4}{3~k T_D} \coth(\frac{T_D}{2T}))$.

On surfaces and along the surface normal (i.e. along $z$), the local harmonic oscillator strength is expected to be half that of the bulk due to the absence of any layer on top. 
The equipartition of energy is accounted for by considering a surface Debye temperature $T_{Ds} \sim T_D/\sqrt(2)$. 

\subsection{Different scattering regimes}

\begin{figure}	\includegraphics[width=0.8\linewidth]{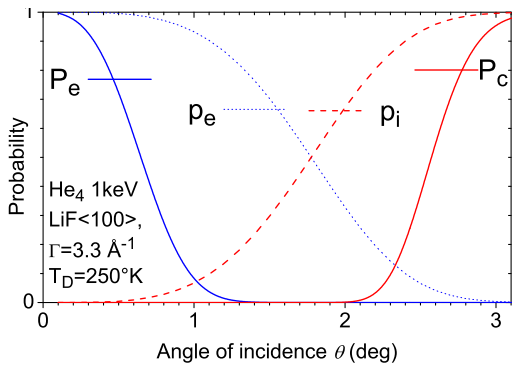}
	\caption{(Color online) The overall purely elastic probability $P_e = DWF$  (blue full line) is evaluated as a function of the angle of incidence. The equivalent colliders model allows derivation of the individual elastic  $p_e=P_e^{1/N}$ (blue dashed), inelastic probability $p_i=1-p_e$ (red dashed) and the overall classical probability $P_c=p_i^N$ (red full line). The quantum and classical regime appear separated by a broad mixed regime.
		\label{fig_Pclassique}}
\end{figure}

The quantum and classical regimes are often identified by the presence or absence of diffraction features\cite{Rousseau_2007,Schuller_2007}. Here we will consider the quantum regime as defined by the elastic scattering which is a more strict requirement. In the above approach it means that all individual collisions are elastic, and the scattering profile is a delta function at the specular angle, without energy loss associated. Surprisingly the classical limit is more difficult to define. Strictly speaking, the probability $P_c$ that all collisions along the trajectory are inelastic will always be zero. This is because, at comparatively large distances from the surface, in the wings of the profile in Fig. \ref{fig_dEdx}, the elastic probability is unity. By construction, such events are not taken into account in the $N_{eq}$ equivalent scatterer model which considers only the collisions participating actively, so that $P_c$ is well defined as $(1-p_e)^{N_{eq}}$. This probability is displayed in Fig.\ref{fig_Pclassique} for 1 keV helium atoms. It shows that the quasi-elastic and quasi-classical regimes are completely separated and that a mixed regime is present in between. 
Here, the predictions of the purely classical behavior would be only slightly overestimated due to the lack of contribution from the significant number of elastic collisions.
Before discussing the associated energy loss distribution and angular scattering profiles, the effect of a single inelastic collision is investigated in detail.

\section{A single inelastic collision}

Taking into account all possible inelastic transitions at a finite temperature is complex in quantum mechanics, even for a harmonic oscillator,  whereas it is comparatively simple using classical mechanics. One simply considers position and momentum distributions given by the Gaussian quantum probability in Eq.\ref{z2T}. In the present case, the collision time of keV projectiles is smaller than the typical vibration period by several orders of magnitude. Compared with TEAS the interaction time with the surface is the same ($\tau=1/\Gamma v_\perp$) but it is typically $N_{eq}$ times smaller with each surface atom. The sudden approximation corresponds here to the situation of an incident atom interacting with a frozen surface. Only the position distribution has to be taken into account in an inelastic collision by randomly distributing the surface atoms around their equilibrium position (see e.g.\cite{Winter_2002,Villette_2000,Pfandzelter}). 

\subsection{Inelastic angular and energy profiles}
At a distance $z$ from the surface, the projectile deflection, associated with a surface atom at its equilibrium position, corresponds to the elastic value $\theta_e \propto  e^{-\Gamma z}$ (height of the $\gamma_z$ peaks in Fig.\ref{fig_gamma}). If the surface atom is displaced below or above its equilibrium value by $\sigma_z$, the actual impact parameter $z$ becomes $z\pm\sigma_z$ and the deflection is distributed around the elastic reference $\delta \theta_\pm =\delta\theta_e e^{\mp\Gamma \sigma_z}$ \textit{i.e.} an angular distribution which is the exponential of a normal position distribution. Such a distribution is known as the log-normal distribution; $P(\delta\theta)=LN[\delta\theta_e;\Gamma\sigma_z](\delta\theta)$. This distribution $LN[x_0;w](x)$ is characterized by its median value $x_0$ and a scale parameter $w$. Here the median value is the elastic scattering angle $x_0=\delta\theta_e$ and the scale parameter is $w=\Gamma\sigma_z$ with $\sigma_z^2$ the variance of the normal distribution and $\Gamma$ the coefficient in the exponential form linking $\delta\theta$ and $z$

\begin{equation}
LN[x_0;w](x)=\frac{A}{\sqrt{2\pi}wx}exp(\frac{-(\ln \frac{x}{x_0})^2}{2w^2} )  \label{LN} \end{equation}

The scattering $\delta\theta$ angle appears as a ratio to the median value $\delta\theta_e$ associated with the equilibrium position $z=0$. 
For the simple interaction potential considered here, $\Gamma$ is fixed and the scale parameter $w=\Gamma\sigma_z$ does not depend on the impact parameter (turning point) $z_0$. 
The scale parameter $w$ is therefore the same for all binary collisions along the trajectory and whatever the angle of incidence $\theta$ providing a universal angular profile for individual deflection where only the magnitude varies. 
The width $\sigma_\theta$ of this profile is proportional to the deflection angle and can be defined via the variance $\sigma_{\theta}^2$ of the log-normal distribution. 
\begin{equation} \sigma^2_\theta=e^{w^2}(e^{w^2}-1)\delta\theta_e^2   \label{sigma_LN} \end{equation}
 
The variance of the angular broadening induced by an individual inelastic collision is therefore proportional to the small angle binary recoil energy ($E_r=\mu E \delta\theta_e^2$) \textit{i.e.} $\sigma_{\theta}^2=e^{w^2}(e^{w^2}-1)\frac{E_r}{\mu E}$. 
 
The recoil energy $E_r$ reported here is only the central recoil energy associated with an inelastic collision $E_r=\mu E \delta\theta_e^2$. 
The energy loss profile of the projectile can be obtained by considering that the values $\delta E_{\pm}$ associated with a displacement of the surface atoms by $\pm\sigma$ are $\delta E_{\pm} = E_r~e^{\mp2\Gamma \sigma}$. 
This leads to a log-normal distribution $P(\delta E)=LN[E_r;2\Gamma \sigma_z](\delta E)$ of the energy loss with a scale parameter $w=2\Gamma\sigma_z$ i.e. twice the  width of the angular deflection distribution, due to the quadratic dependence of the energy loss on the angular deflection. 
The inelastic angular profile is considered as a broadening around the elastic value $\delta\theta$. The energy profiles are different since, by definition, the elastic scattering does exchange energy and is therefore not centered around $E_r$. This is consistent with the fact that for an elastic collision the wave function is left unchanged in Eq.\ref{DWF_overlap}.

\subsection{Out of plane broadening}
In the previous sections, the scattering was described only in the specular plane (along $z$), either with the planar surface model or for trajectories located on top of a row of atoms. 
Within these "top row" trajectories, the out-of-plane inelastic deflection originates from a target displacement inside the surface plane and perpendicular to the specular $(x,z)$ plane,\textit{ i.e.} along the $y$ direction. A position $\delta y$ of the scattering center will induce a lateral deviation $\delta\theta_y$. This corresponds in Fig.\ref{fig_gamma} to a surface atom displaced out of the figure plane and producing a rotation by $\delta\theta_y$ of the scattering plane. This position $\delta y$ is normally distributed with a variance $\sigma_y^2$ determined by the bulk Debye temperature \textit{i.e.} $\sim$ half of $\sigma_z^2$.
The distance to the target is now $\rho=\sqrt{z_0^2 + y^2}$ and the scattering plane is tilted by an angle $\alpha=\arctan(y/z_0)$. The deviation $\delta\theta_y$ is $\delta\theta_y = \delta\theta_e \sin\alpha$. 

For perpendicular energies $E_\perp \lesssim 1eV$, $z_0\gtrsim$ few $\AA$ so that for reasonable surface temperature, the ratio $\sigma_y/z_0\lesssim 1/10$ suggesting further simplifications of $\rho\simeq z_0$ and $\sin(\alpha)\sim \alpha$. This leads to a linear form  $\delta\theta_y = y~\delta\theta_e/z_0$ indicating that, at this position, the typical lateral deviation $\delta\theta_y$ is an order of magnitude smaller than $\delta\theta_e$ and that $\delta\theta_y$ should follow a normal distribution if the $z$ variation is neglected

\begin{equation} \delta\theta_y = \frac{\Gamma a\theta^2~y}{3 z_0},\sigma_{\theta_y} =\frac{\Gamma a\theta^2~\sigma_y}{3 z_0} \label{eq_phi} \end{equation}

\subsection{Averaging over the lattice unit} \label{sec:averaging}

\begin{figure}	\includegraphics[width=0.9\linewidth]{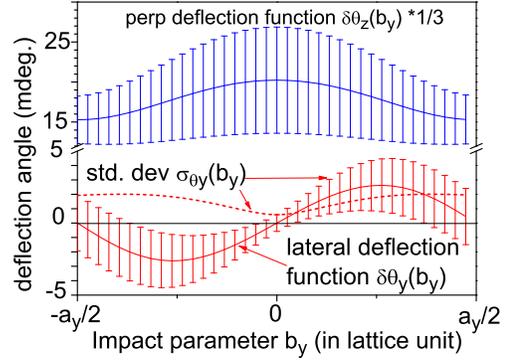}
	\caption{For 1keV helium atoms along the $<100>$ direction, the elementary elastic deflection functions $\delta\theta_z(b_y)$ (top) and $\delta\theta_y(b_y)$ (bottom) are plotted with the inelastic broadening $\sigma_{\theta_z}$ and $\sigma_{\theta_y}$ obtained by distributing the scattering center located at $b_y=0$ by a thermal gaussian distribution with $\sigma_z$ from Eq.\ref{z2T} and $\sigma_y=\sigma_z/\sqrt{2}$.\label{fig_deflection_function}}
\end{figure}

Eq.\ref{eq_phi} discussed above indicates that the on-top situation is not representative of lateral momentum transfer mainly because the angle $\alpha$ of the scattering plane is centered around zero whereas this angle $\alpha$ can be significant for positions $y$ close to that producing the rainbow scattering angle\cite{Winter_rep_2002} i.e. such that $d\theta_y / dy=0$. 

The actual profiles of the momentum transfer both in the specular plane ($\delta\theta$) and perpendicular ($\delta\theta_y$) have to be evaluated over all possible impact parameters forcing us to abandon the planar model and adopt the string model \cite{danailov,danailov_2007}. In the string model, or row model, the integration of the individual binary contribution is performed along the identical rows. Along the $<100>$ direction only one string (a row of alternating $F$ and $Li$) is needed per lattice site (inset in Fig.\ref{fig_dEdx}) 
\begin{equation} V_s(\rho)= 2V n_x K_0 (\Gamma\rho),~\rho=\sqrt{y^2 + z^2}\label{danailov} \end{equation}
Where $n_x$ is the linear density and $K_0$ is the modified Bessel function of the second kind.
Close to the surface ($z<a$), the potential energy landscape can be estimated accurately by summing the contributions of only five rows, a central one and two adjacent rows on either side. The turning point $z_0$ now depends on the lateral impact parameter $b_y$ defining the corrugation function $z_c (b_y)$.
In this description, the elastic contribution is given by deflection functions $\delta\theta_z(b_y)$ along $z$ and $\delta\theta_y(b_y)$ along $y$ of an individual elastic collision with all surface atoms still at their equilibrium position. 

The corresponding inelastic scattering profile is now derived by distributing the central atom according to the $\sigma_y$ and $\sigma_z$. Each point of the elastic deflection function becomes a distribution and we follow these $\delta\theta_z(b_y)$ and $\delta\theta_y(b_y)$ distributions along the corrugation function. 
Taking the corrugation function $z_c(b_y)$ corresponding to $E_\perp = 0.1 eV$, the values of the elastic scattering along $y$ and $z$ and their standard deviations  $\sigma_{\theta_z}$ and $\sigma_{\theta_y}$ are plotted in Fig.\ref{fig_deflection_function} for the $<100>$ direction. In this direction, the linear periodicity within the string is $a\sqrt{2}$  while the string periodicity or transverse periodicity, as observed in diffraction, is $a_y=a/\sqrt{2}=2.015\AA$. 

\begin{figure}	\includegraphics[width=0.85\linewidth]{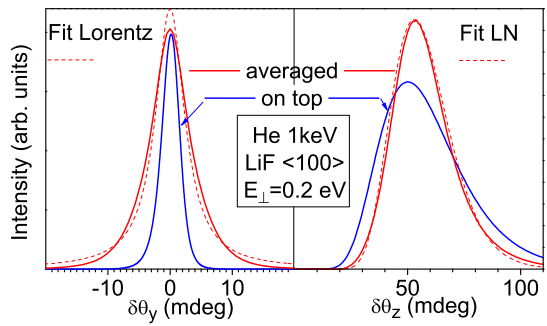}
	\caption{Angular straggling $\delta\theta_z$ (right) and $\delta\theta_y$ (left) of an individual inelastic collision averaged over the lattice cell. Compared with the "on top" trajectory, the lateral broadening has acquired a Lorentzian character with  $w_L\sim 3$ times that of  Eq.\ref{eq_phi} while the log-normal scale parameter $w_z$ describing the broadening of the polar distribution is reduced by $\sim 30\%$.	\label{fig_Lorentz}}
\end{figure}

As anticipated, $\sigma_{\theta_y}$ is minimum on top of a row (the scattering plane is perpendicular to the displacement), corresponding in Fig.\ref{fig_deflection_function} to  $b_y=0$ and almost $\sim 3.5$ times larger in the bottom of the corrugation function where the displacement has a large projection into the scattering plane. 
In addition, the shape of the inelastic scattering profile averaged over the lattice unit are displayed in Fig.\ref{fig_Lorentz}. The  $\delta{\theta_z}$ inelastic angular distribution, along the $z$ axis, still shows a pronounced log-normal character but clear departures can be observed for the tails. The inelastic width $\sigma_{\theta_y}$ at $b_y = \pm a_y /2$ is almost half that at $b_y=0$ resulting in an average scale parameter reduced by $\sim30 \%$. This ratio originates mainly from geometric projection of the $z$ contribution and could be system specific.

Fig.\ref{fig_Lorentz} shows that the inelastic $\delta\theta_y$ distribution is almost gaussian for on-top conditions as in Eq.\ref{eq_phi} but large side wings are produced by the tails of the log-normal distributions on both sloping sides of the corrugation function. 
In this geometry, both the displacements in $y$ and $z$ contribute to the $\delta\theta_y$ profile. 
A lorentzian profile is superposed showing a resemblance but also clear departure on the wings. The standard deviation averaged over the lattice unit is almost three times larger than the prediction of Eq.\ref{eq_phi} which was restricted to on-top trajectories. 

The inelastic angular width $\sigma_{\theta_y}$ depends both on the in-plane and out-of-plane movement $\sigma_y$ and $\sigma_z$. 
Since these two values are proportional to each other, the ratio of $\sigma_{\theta_y}$ to $\sigma_{\theta_z}$ should not depend on temperature. 
Also, the ratio should hardly evolve with the angle of incidence because the turning point $z_0$ varies smoothly so that the geometry changes very slowly. 
Last but not least, the integration over the lattice unit shows that, for moderate angles of incidence, the length of the trajectory does not vary significantly but the associated energy loss does. This is visible in Fig.\ref{fig_dEdx} where the energy deposition curves associated with 'on-top' and 'in-between-rows' trajectories both display a width comparable to that of the planar model, but with quite different magnitudes.

\section{the classical limit}
\subsection{angular and energy profiles}

\begin{figure}	\includegraphics[width=0.8\linewidth]{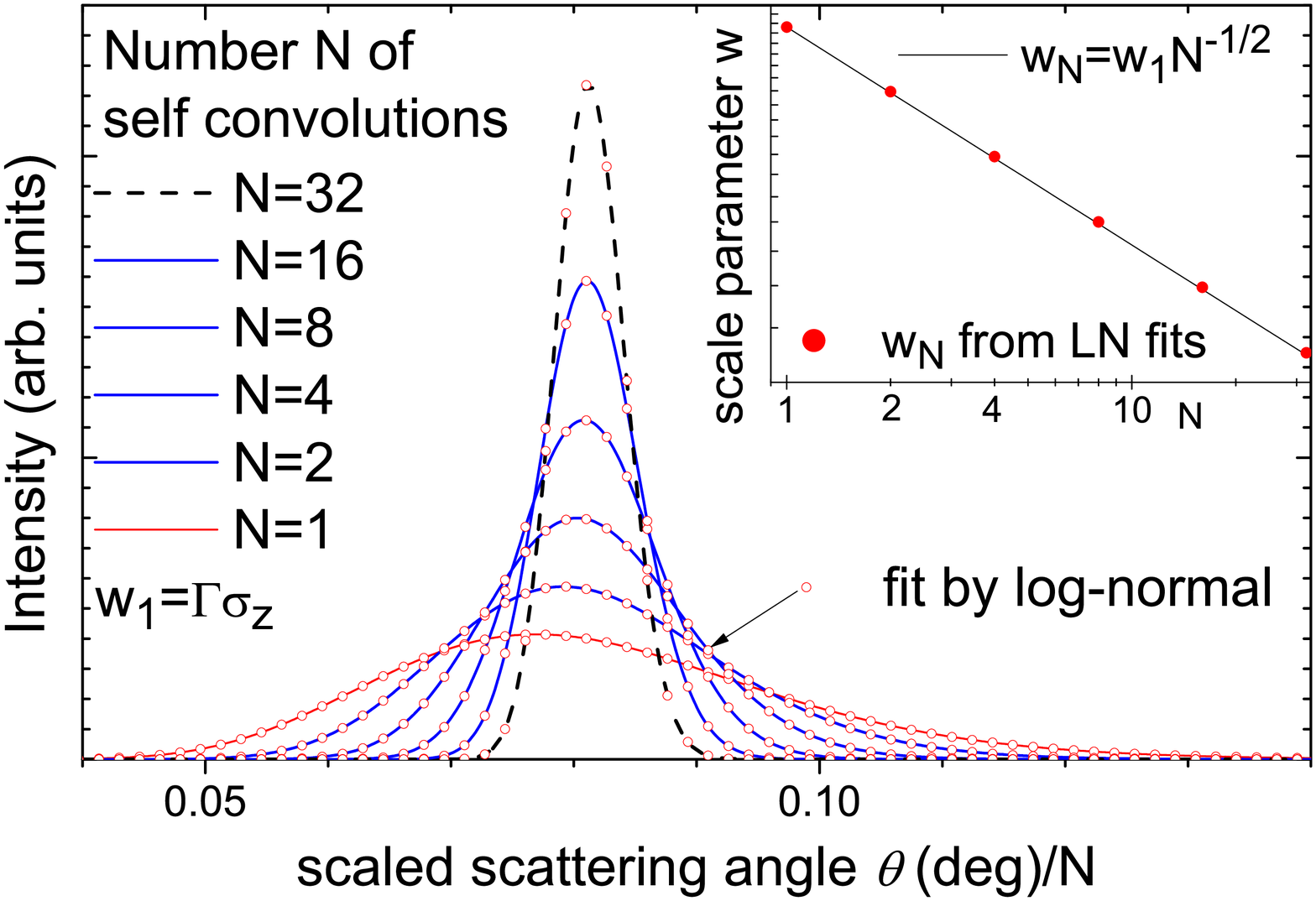}
	\caption{(Color online) The log-normal angular profile of a single scattering event $LN_1(\theta)$ (red curve) is self convoluted N times (blues curves). They are displayed using a 1/N scale. These are well fitted by log-normal distributions (red dots) having a log-normal scale parameter $w_N=w_1/\sqrt{N}$ (inset).	\label{fig_LN_conv}}
\end{figure}

As stated above it is not possible to reach a condition where all collisions would be inelastic. There will always be a significant probability that collisions on the wings of the energy deposition curves are elastic. 
The classical angular distribution is defined here as the one corresponding to an energy loss equal to the classical limit derived in the planar model in Eq.\ref{Eloss}.
The resulting angular variance $\sigma^2_{cl}$ will be the sum of individual variance. As each individual variance is linked to the associated recoil energy, the resulting variance is given by the sum of the recoil energies $E_{loss}=\Sigma E_r$ in Eq.\ref{Eloss}

\begin{equation} \label{sigma_cl}
\begin{aligned}
&\sigma^2_{cl}=e^{w^2}(e^{w^2}-1)\frac{E_{loss}}{\mu E},~w=\Gamma \sigma_z \\
&\sigma^2_{cl}=e^{w^2}(e^{w^2}-1)\frac{2}{3} \Gamma a \theta_{in}^3  
\end{aligned}
\end{equation}

where $E_{loss}$ from Eq.\ref{Eloss} corresponds here to the measured energy loss.	
This result can be derived also from the $N_{eq}$ model where the ensemble of participating sites is finite and restricted to the $N_{eq}$ most important collisions, each associated with a log-normal scattering profile. The convolutions of log-normal distributions are not log-normal distributions but, probably because here $w<1$, they display a very strong log-normal character as can be seen in Fig. \ref{fig_LN_conv} where successive self convolutions perfectly superimpose with their fit by log-normal distribution with scale parameter $w_{N_{eq}}=w/\sqrt{N}$. 
Both the total energy loss approach and the $N_{eq}$ approach agree on a classical angular distribution corresponding to a log-normal distribution with median value $2\theta$ and a scale parameter $w_{cl}=\Gamma \sigma_z /\sqrt{_{N_{eq}}}$. 

\subsection{Angular and energy correlations}

At the single collision level, the energy and angular distribution are strictly correlated as $E_r=\mu E \delta\theta^2$. Naively, there is a concern that successive convolutions would blur this correlation but this is not the case. The correlation is well preserved so that, for a given angle of incidence $\theta_{in}$, the energy loss depends on the scattering angle $\theta_{out}$ within the angular profile. More precisely it evolves with the cube of the scattering angle referred to the primary beam direction $\Delta E\propto(\theta_{in}+\theta_{out})^3$.

\subsection{Neon LiF}
Only few experiments have measured the energy loss for neutral projectiles at grazing angle of incidence. The main reason is that electrostatic analysis is not possible and that sub-eV accuracy is desirable on top of keV energies. To our knowledge, only Neon atoms have been carefully investigated between 1 and 5 keV on LiF target oriented along a random direction and for angle of incidence larger than one degree\cite{Villette_these,Pfandzelter,Mertens_2000,Winter_2002}. 
According to Eq.\ref{Pes} this corresponds to a situation close to the classical limit. 

J. Villette\cite{Villette_these} showed that the energy loss profile can be well described by a log-normal distribution, where the scale parameter $w$ varies slowly with the angle of incidence and depends only on the surface temperature for fixed incidence angles. The measured energy loss $\Delta E$ was found to depend both on the angle of incidence and on the outgoing angle ; it was found proportional to the overall scattering angle $\Delta E \propto (\theta_{in} + \theta_{out})^3$. 
For grazing incidence data using Ne atoms between 1-3 keV on a LiF surface at room temperature, all the data could be described by $\frac{\Delta E}{E}= \alpha (\theta_{in} + \theta_{out})^3$ with $\alpha=8\pm3~10^{-6}$ if $\theta$ is expressed in deg. 
The log-normal angular and energy profiles as well as the cubic dependence were reasonably well reproduced in numerical simulations using randomly displaced surface atoms (see Eq.\ref{z2T}) with a surface Debye temperature of 539°K. From the simulations it was possible to derive the number of surface atoms which participate actively in the scattering process, using a range $\Gamma=3.5\AA^{-1}$ from Ref\cite{Villette_2000}. 

Soon after, a planar model was developed to make the link between the log-normal scale parameter $w$ and the range $\Gamma$ of the interaction potential\cite{kristel,Manson2008}.
Similar qualitative conclusions were reached by \cite{Pfandzelter,Mertens_2000,Winter_2002} with higher projectile energy and a surface temperature twice as large. 
These authors also developed an elaborate tracking of systematic errors and came up with a value of $\alpha$ almost twice larger. They suggested a surface Debye temperature of 250K instead of 539K.

All these observations find a natural interpretation in the present frozen lattice formalism where the shape and correlations of these quantities are calculated and linked together without adjustable parameters. In the present form, using the range parameter of the binary interaction potential in\cite{Villette_these,Pfandzelter} we obtain $\frac{\Delta E}{E}=\frac{2\mu\Gamma a}{3\times 8}\sim 4~10^{-6} (\theta_{in} + \theta_{out})^3$, which indicates that the planar model is capable of semi quantitative prediction. 
Eq.\ref{Eloss} indicates an energy loss independent of the temperature, but a surface Debye temperature as low as 250K means larger amplitudes of surface atoms. For instance, the $Li^+$ ions would not be completely hidden by the $F^-$ ions as indicated in Fig.\ref{fig_gamma}. For an equivalent momentum transfer, the recoil energy of these $Li$ ions is three times larger due to their lighter mass. More simulation work is needed to take into account the contribution of different species in the surface unit cell. 

\section{mixed quantum-classical regime}

This is the regime where both $p_e^N$ and $(1-p_e)^N$, the probabilities for the successive collisions to be all elastic or inelastic respectively, are far from one (red and blue curves in Fig.\ref{fig_Pclassique}). The observables such as the energy loss and angular profiles should lie in between the delta function of the quantum regime and the broader log-normal distribution discussed above.
The actual mean energy loss results from the Lamb-Dicke weighting of all individual collisions along the trajectory \textit{i.e.} $\Delta E=\Sigma_{i=-N}^{i=+N} \delta E (i) P(\delta E)$ with $i=x/a$ and $P(\delta E (x))$ given by Eq.\ref{dEt}. 
In contrast to $E_{loss}$ which was defined earlier as the sum of the possible (virtual) recoil energies, (becoming real in the classical limit), $\Delta E$ is the actual energy loss \textit{i.e.} the sum of the inelastic events. 
The mean variance of the inelastic angular profile is 
\begin{equation} \label{sigma_ine_moy}
\sigma_{ine}^2 = e^{w^2} (e^{w^2} -1) \frac{\Delta E}{\mu E }, w=\Gamma \sigma_z
\end{equation}
This mean variance lies well below the classical limit $\sigma_{Cl}$ of Eq.\ref{sigma_cl} (as displayed a little further in Fig. \ref{fig_sig_polar}). The curve starts with a linear behavior (see Eq.\ref{sigma_sc} below) and then merges with the $E_{\perp}^{3/4}$ classical dependence, implicit in Eq.\ref{sigma_cl}.  

Alternately, given the (quasi gaussian) energy deposition profile displayed in Fig.\ref{fig_dEdx}, the statistical weight of any combination of elastic and inelastic collision can be calculated to generate the proper combination of all the ($\sim N_{eq}!$) associated scattering profiles, instead of using the one associated with the average energy loss. 
In addition, the development in perturbation can be expanded in terms of the number $N_{ine}$ of inelastic events where all contributing profiles weighted by their probability are taken into account.
For illustration purposes, this can be done by hand within the $N_{eq}$ model, keeping in mind that the flat probability distribution is a poor representation of the quasi gaussian one displayed in Fig.\ref{fig_dEdx}.

The equivalent colliders model assumes independent events with well defined probability $p_e$ and $p_{ine}=1-p_e$. The angular profile $P_{ine}(\theta)$ can thus be cast in a binomial form where the number of inelastic collision $N_{ine}$ among $N_{eq}$ is reflected in the binomial coefficient $\binom {N_{eq}} {N_{ine}}$

\begin{equation*} 
\begin{aligned}
&P(N_{ine}) =~\binom {N_{eq}} {N_{ine}}~p_e~^{N_{eq}-N_{ine}} (1-p_e)^{N_{ine}}  \\
&P_{ine}(\theta) =\Sigma^{N_{eq}}_{N_{ine}=1}~P(N_{ine}) LN_{N_{ine}} (\theta) \nonumber 
\end{aligned}
\end{equation*}

\begin{figure}	\includegraphics[width=0.8\linewidth]{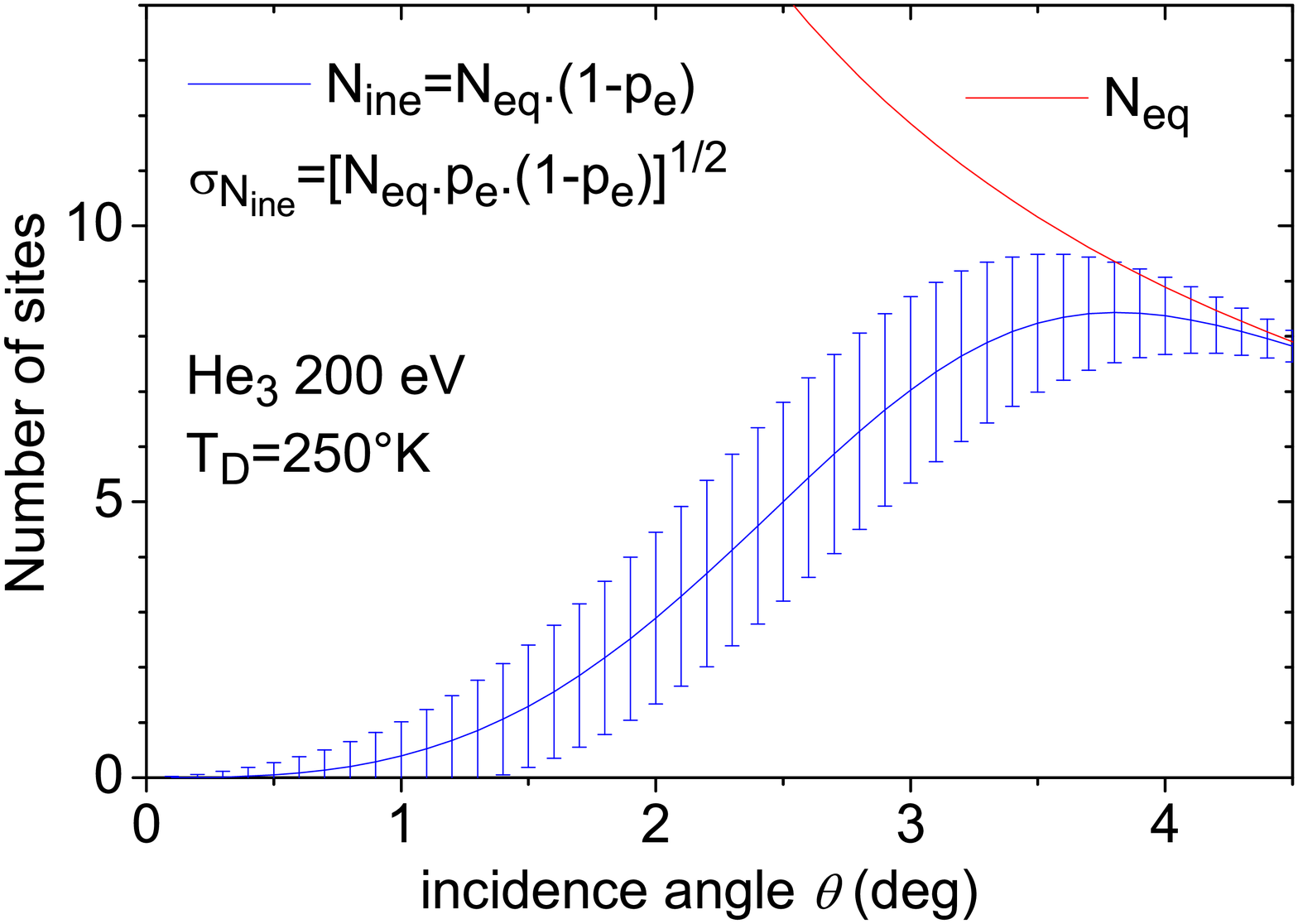}
	\caption{(Color online) For 1 keV He atoms, the red curve indicates the number of lattice sites participating to the deflection (trajectory length) as a function of $\theta_{in}$. The blue curve indicates the number $N_{ine}$ of inelastic collisions  that actually participate to the energy loss and angular straggling.	\label{fig_Nin}}
  \end{figure}

For large $N_{eq}$ this distribution is characterized by its mean value $\langle N_{ine} \rangle=(1-p_e)~N_{eq}$, and variance  $N_{eq}~p_e~(1-p_e)$. These quantities are plotted in blue in Fig.\ref{fig_Nin} with the standard deviation as an error bar. The mean angular straggling and mean energy loss will simply be given by the classical value (fully inelastic) multiplied by $p_{ine}=1-p_e$, the blue line in Fig.\ref{fig_Nin}. This line indicates approximately how the observables connect to the classical behavior.

The consequence on the scattering profiles are quite significant since the final variance is only $N_{ine}$ times that of a single collision, much less than the $N_{eq}$ of the classical profile.
For a number of  inelastic collisions exactly $N_{ine}$ (among $N_{eq}$), the scale parameter is $w_{N_{ine}} =w \sqrt{N_{in} }/N_{eq}$. The mean scattering profile corresponds to a scale parameter $w_{mean}=w_{Cl}\sqrt{(1-p_e)}$ which can be much narrower than the classical limit, the latter being much narrower than the individual 
scattering width (central and external curves in Fig.\ref{fig_LN_conv}).


Returning to the more realistic planar model, the energy deposition curve is more localized and so is the inelastic probability distribution. In the quasi-elastic regime, \textit{i.e.} when Eq. \ref{Pes} gives an overall probability larger than few percent, the individual inelastic probabilities (Eq.\ref{pe}) are small enough to be approximated by $p_{ine}=1-p_e=1-e^{-\beta\delta E}\simeq \beta\delta E$ with $\beta= \frac{3}{k T_D} \coth(\frac{T_D}{2T})$. 

The probability follows the same gaussian like distribution so that the weighted distribution should follow an even more localized distribution with a variance reduced by a factor two. This suggests that the inelastic properties will be governed by the few central collisions making the $N_{eq}$ model inappropriate in this quasi-elastic regime where only one or two inelastic events contribute to the inelastic profile.
The most probable angular broadening will be associated with $\delta E_{max}$, the peak of the energy deposition curves (Eq.\ref{DE_Max}),  giving a standard deviation of the $\theta_z$ inelastic angular profile (Eq.\ref{sigma_LN}) 

\begin{equation}\label{sigma_sc} 
\begin{aligned}
&\sigma_{sc}^2 = e^{w^2} (e^{w^2} -1) ~\Gamma^2 a^2 \theta^4 /4\\
&\sigma_{sc} =  \Gamma a \frac{E_\perp}{2E}  ~(e^{~\Gamma^2 \sigma_z^2} (e^{~\Gamma^2 \sigma_z^2} -1))^{1/2}\\
\end{aligned}
\end{equation}
which should be characteristic of the quasi-elastic regime. 

The $\theta^2$ dependence indicates that the inelastic width can be extremely small at the lowest grazing angles and Eq.\ref{sigma_sc} can be useful to point out the angular resolution needed to resolve inelastic events. $\sigma_{sc}$ can also be expressed as 
$\sigma^2_{sc}\sim \sigma^2_{cl} \frac{3\Gamma a \theta_{in}}{8} $. 
This is approximately twice as large than predicted by the $N_{eq}$ model which also gives a linear behavior but with $\sigma^2_{sc}\sim \sigma^2_{cl} /N_{eq}=\sigma^2_{cl}\frac{\Gamma a \theta_{in}}{6}$ because the average value considered in the $N_{eq}$ model is $\sim$ half the value of the peak.

\subsection{Temperature dependence}
Two temperatures are present in the model. The surface Debye temperature $T_D$ describes the most important surface property here, namely the frequency of the Debye local oscillator. The Debye frequency, when expressed as a temperature, gives an idea whether, at a given temperature $T$ the surface atoms are mainly in the vibrational ground state or not. 
The Debye temperature enters in two places to calculate the elastic probability. One is via the simple ratio $\frac{E_{loss}}{kT_D}$ and the other is in the term $\coth(\frac{T_D}{2T})$ which also scales as $T/T_D$ providing a high $1/T_D^2$ sensitivity inside the exponent of the elastic ratio as illustrated in Fig.\ref{fig_Pe_200eV}.
The actual temperature $T$ does not enter in the energy loss $E_{loss}$ because the momentum transfer is calculated with respect to the center of the wave-function and is therefore temperature independent. 
The temperature $T$ enters only in the term $\coth(\frac{T_D}{2T})$ and in the inelastic properties.
It determines the spatial extent $\sigma_z$ of the surface atoms (see Eq.\ref{z2T}) and therefore the width $w=\Gamma\sigma_z$ of the log-normal scattering profile of an individual inelastic event.
For small values of the scale parameter ($w^2 \ll 1$), the pre-factor present in the variance of the log-normal distribution can be simplified $e^{w^2}(e^{w^2}-1)\approxeq w^2$ so that the quasi elastic angular width (Eq.\ref{sigma_sc}) receives a compact form.
\begin{equation}\label{sigma_sc_lin} 
  \sigma_{sc} \sim  \Gamma a \Gamma \sigma_z \frac{E_\perp}{2E}.\end{equation} 
In this respect the He-LiF system is probably not a favorable case because large value of the work function usually mean large value of $\Gamma$ and light mass of surface atoms contribute to large values of $\sigma_z$.

\section{Inelastic diffraction}

The inelastic processes have been described as angular straggling around the elastic scattering values. Each deflection is so tiny that it hardly perturbs the overall trajectory but induces significant broadening in the final angle. 
The associated recoil energy is, on average, less than a vibration quantum and does not allow which-path information that would prevent diffraction. 
Of course inelastic events with different final momentum do not interfere with the elastic one, even if some inelastic event can accidentally end up exactly at Bragg position. This makes the line profile and data analysis more complex. 

In the following sections, we use existing data\cite{DebiossacPRL,Rousseau_2007} on the well investigated helium LiF system for which the parameters of the model binary interaction potential have been published\cite{DebiossacPRL}. The direction $<100>$ (inset in Fig.\ref{fig_Ctp_vs_Ep}) was chosen because only one row of alternating Li and F ions is needed with a negligible role from the Li ions at limited temperature and perpendicular energy $E_\perp$ (see Fig.\ref{fig_gamma}). 
Before comparing with the predictions of the model developed in the previous sections, we briefly review some of the specific aspects of inelastic diffraction in the experiments.

\subsection{Data (re)analysis}
 
\begin{figure}	\includegraphics[width=0.90\linewidth]{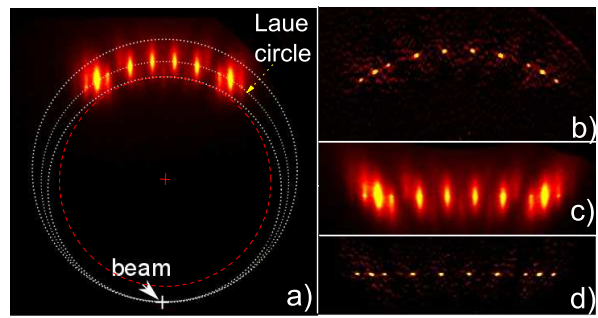}
	\caption{{For 460 eV He$_4$ at $1.57^\circ$ (\textit{i.e.} $E_\perp=345$ meV comparable to Ref\cite{Aigner_2008,Seifert_2015}), the diffraction circles containing the beam position in the raw image a) are transformed into horizontal lines in c). The effect of the doubly differential filter applied in the $z$ direction and isolating the elastic $k_y$ profile and intensity\cite{Debiossac_Nim_2016} is illustrated in b) and d), the $k_z$, vertical extension is then given by the bandwidth of the filter.}}
	\label{circles}
\end{figure}

In the early experiments where well resolved diffraction features were first observed with fast atoms\cite{Schuller_2007,Rousseau_2007}, there was no clear evidence of an increased intensity at the Laue circle (see e.g. Fig.\ref{circles}).  
In this context where the energy is not conserved, even the central concept of wavelength is not uniquely defined .
The detector is located far away from the surface, so each pixel corresponds to well defined scattering angles $\theta_y,\theta_z$ or $k_y,k_z$ with the fast motion $k_x$ being perpendicular to the detector. 

\textit{A priori}, two polar transformations are possible to associate these scattering angle to a diffraction circle ($k_{eff}$) while preserving the coordinate $k_y$ where diffraction is observed. One taking the center of the Laue circle as a universal reference \textit{i.e}. the shadow edge (red circle in Fig.\ref{circles} concentric to Laue circle), and the other one referring all angles to the position of the direct beam (white circles in Fig.\ref{circles}). 

Schematically the scattering by the surface consists of an incoming and an outgoing part.
The first polar transform considers that only the part leaving the surface is important while the incoming trajectory is forgotten.
The second one illustrated in Fig.\ref{circles} considers, for each pixel, a diffraction circle intersecting to the primary beam and having a diameter that is the average between incoming and outgoing trajectories. 

In the previous discussion, the total scattering angle appears in several equations as a natural reference for all trajectories. 
The optical analogy implicit in the HCW model allows a simple interpretation. In the HCW model, the surface corrugation function is a mirror with a shape $z_c(y)$ and the phase difference responsible for the diffracted intensity is $2 k_{in} z_c$ in the specular plane.
The inelastic diffraction can be regarded as a modification of the wavelength by the surface. The phase difference between possible paths naturally splits into two terms corresponding to the incoming and outgoing wave-vectors resulting in phase modulation $(k_{in}+k_{out}) z_c$ with  $k_{eff}=(k_{in}+k_{out})/2$.
The relevant circle contains the primary beam and the pixel of interest and is forced to preserve the specular plane. With this transformation an effective wave vector $k_{eff}$ is associated to any pixel on the detector\cite{Debiossac_Nim_2016}. 
All circles become horizontal lines while the diffraction coordinate $k_y$ is left unchanged showing evenly spaced diffraction peaks $k_y = m G_y$ labeled according to the specular one. In elastic diffraction, these intensities $I_m$ are directly connected to the form factor, \textit{i.e.} to the scattering elements inside the unit cell. In the present case the form factor is the potential energy landscape of the frozen lattice unit cell. 
\begin{figure}	\includegraphics[width=0.85\linewidth]{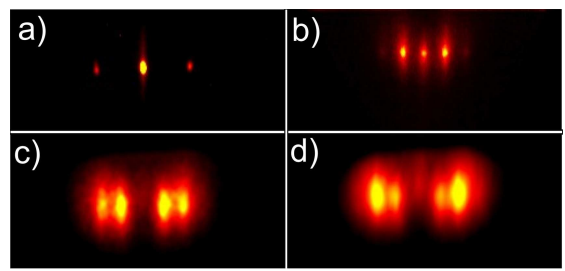}
	\caption{Polar transformed diffraction patterns of He on LiF along the $<100>$ direction. a) 460 eV from Ref\cite{DebiossacPRL} and b) c) d) 200 eV from Ref\cite{Rousseau_2007}. Each horizontal line corresponds to a well defined $k_{eff}$. The polar plots in  Fig.\ref{fig_hi_Im} and Fig.\ref{fig_y05_Py_fit} correspond to projections on the vertical axis\textit{ i.e}. P($k_{eff}$).	\label{fig_Images_diff}}
\end{figure}

\begin{figure}	\includegraphics[width=0.9\linewidth]{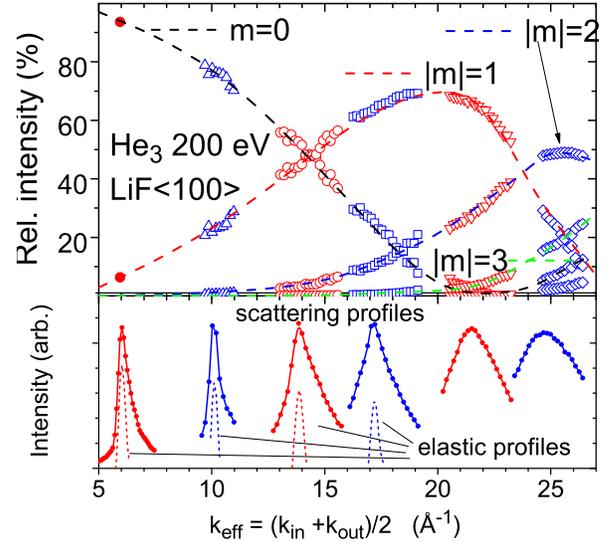}
	\caption{Scattering profiles as a function of the average momentum $k_{eff}$ for six different angle of incidence\cite{Rousseau_2007} (Lower panel). The elastic component is estimated from a doubly differential filter (see Fig.\ref{fig_y05_Py_fit} for a better estimate). The corresponding relative intensities on the top panel show no singularity when passing though the elastic component. 	\label{fig_hi_Im}}
\end{figure}

To analyze how the relative intensities evolve on either side of the Laue circle, we use an automatic fitting procedure using a multi-parameter profile that can adapt to different line-shape but that is common to all diffraction orders. 
For each given value of $k_{eff}$ the fit produces a line-shape and peak intensities $I_m$. 
For successive values of  $k_{eff}$ the fit leaves the line-shape free to evolve independently from previous $k$ values. 
The fitted relative intensities are reported in the top panel of Fig.\ref{fig_hi_Im} for diffraction images recorded with 200 eV $He_3$ projectiles on LiF$<100>$ at six different incidence angles\cite{Rousseau_2007} with their associated profiles reported in the bottom panel of Fig.\ref{fig_hi_Im}. 
Obviously, there is a smooth continuity and, looking only at from the relative diffracted intensities it is impossible to guess the location of the Laue circle. Only the polar profiles indicated below and in Fig.\ref{fig_y05_Py_fit} indicate the location of the Laue circle\cite{soulisse}. 

Here, the elastic diffraction profiles reported as dashed lines under the scattering profiles are obtained by passing a doubly differential 'Mexican hat' filter having a bandwidth equal to that of the primary beam $\sigma_\theta$ on the 2D transformed images\cite{DebiossacPRB_2014,Debiossac_Nim_2016,Debiossac_ASS_2017}. 
It relies on the fact that inelastic profiles evolve more smoothly than the elastic profile with the scattering angle $\theta$. Subtracting the half sum of the intensities located at angle $\theta + \sigma_\theta$ and $\theta - \sigma_\theta$ from the intensity at $\theta$ gives then only the rapidly varying elastic intensity at the Laue circle.
Though not fully quantitative, this procedure\cite{Debiossac_Nim_2016} provides a value of the Laue circle and gives an indication of the absolute elastic fraction. 
These estimated elastic peaks are displayed here only to underline that continuity in the intensity ratios $I_m(k_{eff})$ is not accidentally due to the absence of elastic diffraction. It also shows that elastic diffraction was present in the data\cite{Rousseau_2007,DebiossacPRL} but was not identified as such because there was no model of the inelastic profile.
Most important, the intensity ratios $I_m(k_{eff})$ derived on both side of the Laue circle seem to connect to each other as if elastic or inelastic regime were giving identical results. 
This alone is a clear motivation to better understand the inelastic behavior.
Note that profiles analyzed here have an elastic component indicating that the collisions on the surface take place in the quasi-elastic regime and that the continuity of the \textbf{} intensity ratios holds only for scattering values within the fwhm of each polar profile. Beyond this limited angular range, the inelastic intensity ratio $I_m$ departs from the one measured under the elastic component. 

\begin{figure}	\includegraphics[width=0.9\linewidth]{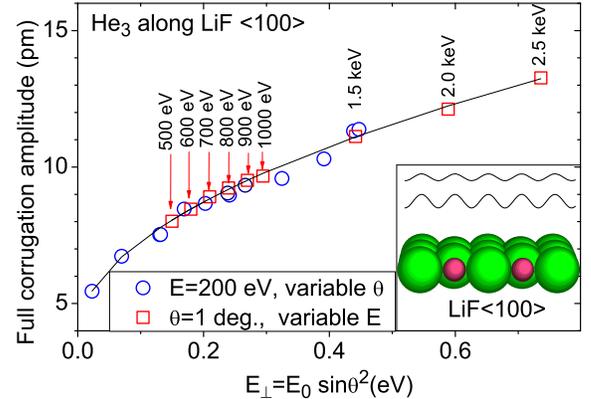}
	\caption{Experimental corrugation amplitude $z_c$ (squares and circles) fitted via a HCW model are reported for different energies and angles\cite{Rousseau_2007,Momeni}. As expected from the axial channeling approximation\cite{Zugarramurdi_2012,Debiossac_PRA,Diaz_2016b}, $z_c$ depends only on the perpendicular energy $E_\perp$. Line is drawn only to guide the eye.	
		\label{fig_Ctp_vs_Ep}}
\end{figure}

To interpret the intensity ratios $I_m(k_{eff})$ we use the HCW model which is here particularly simple along the $<100>$ direction where the LiF corrugation function $Z_c(y)$ was shown\cite{Rousseau_2007,Rousseau_2008,Winter_PSS_2011} to be close to a simple cosine $Z_c(y)=z_c/2 \cos (G_y y)$ with $z_c$ the full corrugation amplitude. 
In this case the HCW predict diffracted intensities $I_m$ given by  $I_m=J_m^2 (2k_{eff}z_c)$ where $J_m$ is the Bessel function of rank $m$. 
A fit to this model allows a direct evaluation of the corrugation function and the results are displayed in Fig.\ref{fig_Ctp_vs_Ep}. 
It indicates clearly that the corrugation amplitude depends only on the perpendicular energy $E_\perp$ (axial channeling approximation) and that, along this $<100>$ direction, $z_c$ increases with $E_\perp$ ; the more the projectile presses on the surface, the larger the corrugation amplitude. This is not surprising here since the minimum of the corrugation function is in between the rows\cite{Momeni}, at a location where there is no atom so that the local repulsion evolves less rapidly that on top of the rows (inset in Fig.\ref{fig_Ctp_vs_Ep}). At larger perpendicular energy ($\ge 10 eV$), the projectile will eventually penetrate in between the rows. Note that the energy region investigated in TEAS is below 100 meV.

\subsection{Elastic ratio}

\begin{figure}	\includegraphics[width=0.95\linewidth]{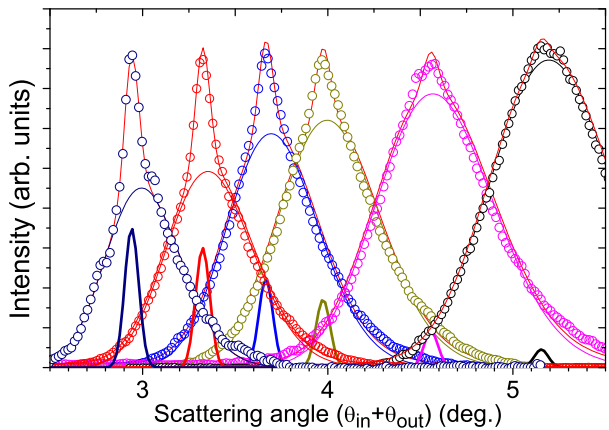}
	\caption{(color online) Polar angle distributions of 200 eV He atoms corresponding to different angle of incidence $\theta_{in}$. The inelastic width and the elastic ratio are estimated using a two component fit. A narrow gaussian component of fixed width $\sigma=0.04$ deg. corresponding to the primary beam profile and a broader a log-normal with free width $w$.	\label{fig_y05_Py_fit}}
\end{figure}

\begin{figure}	\includegraphics[width=0.85\linewidth]{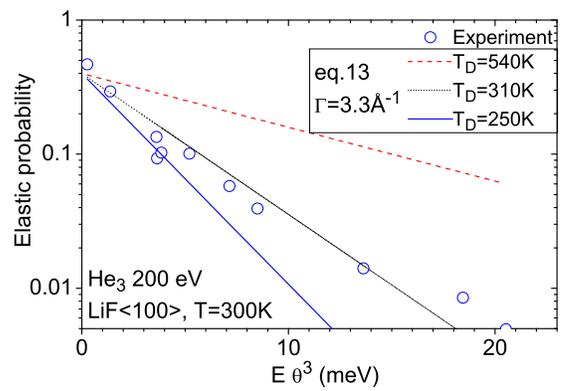}
	\caption{The absolute elastic diffraction probability $DWF$ estimated by the fit of the polar profile in Fig.\ref{fig_y05_Py_fit} are reported as a function of the product $E\theta^3$ and compared with prediction of Eq.\ref{Pes} (scaled by by 0.4) for Debye surface temperatures of 540K\cite{Aigner_2008,Schuller_PRA2010}, 310K\cite{Boato_1976} and 250K\cite{Pfandzelter,Mertens_2000,Winter_2002}. 	\label{fig_Pe_200eV}}
\end{figure}

The DWF or elastic ratio can be estimated from the relative area of the elastic peak. As illustrated in Fig.\ref{circles} the 1D doubly differential filter isolates an almost pure elastic component when applied along $k_z$, \textit{i.e}. perpendicular to the $k_y$ diffraction coordinate. On the Laue circle the resulting 1D profile preserves the relative intensities of the diffraction orders\cite{DebiossacPRB_2014,Debiossac_Nim_2016,Debiossac_ASS_2017} but the absolute intensity is quite sensitive to the bandwidth. 
Here the intensity of the elastic and inelastic components are determined by fitting the polar scattering profile by a gaussian peak with a constant width equal to that of the primary beam profile and a free log-normal profile as illustrated in Fig.\ref{fig_y05_Py_fit}. In this figure, the $e^{-\theta^{3}}\propto 1-\theta^{3}$ attenuation of the elastic ratio of Eq.\ref{Pes} is almost visible with naked eyes.
The height of the elastic peak decreases more or less linearly while both the height and the width of the inelastic profile increase linearly with the polar scattering angle. 
More quantitatively, Fig.\ref{fig_Pe_200eV} displays the absolute elastic fraction determined from the fits in Fig.\ref{fig_y05_Py_fit} as a function of the product $E\theta^3$. 
It shows an exponential decay but with a maximum coherence limited to 50\% and with a slope of $\approx 0.24$ $meV^{-1}$. Assuming a value of $\Gamma$=3.3 $\AA^{-1}$\cite{DebiossacPRB_2014} in agreement with\cite{Schuller_PRA2010}, the results of Eq.\ref{Pes} are reported for quite different values of the surface Debye temperature found in the literature\cite{Aigner_2008,Schuller_PRA2010,Boato_1976,Pfandzelter,Mertens_2000,Winter_2002}. A critical analysis of $T_D$ and, to a minor extend $\Gamma$ is beyond the scope of the paper but the physical assumptions behind the derivation of these numbers will probably have to be investigated in more detail.
The comparison shows that at least 50\% of the decoherence is not accounted for by the present model.
The possible origin will be discussed with help of the polar and transverse inelastic angular profiles.

\subsection{Polar angle inelastic line profiles}

To our knowledge, the shape of the polar inelastic profiles (along $k_z$) has never been analyzed in the diffraction regime.
Even in the quantum monte-carlo description of the decoherence in Ref\cite{Aigner_2008,Schuller_PRA2010}, the $k_z$ profile is reproduced by artificially broadening the projectile wave packet.

\begin{figure}	\includegraphics[width=0.85\linewidth]{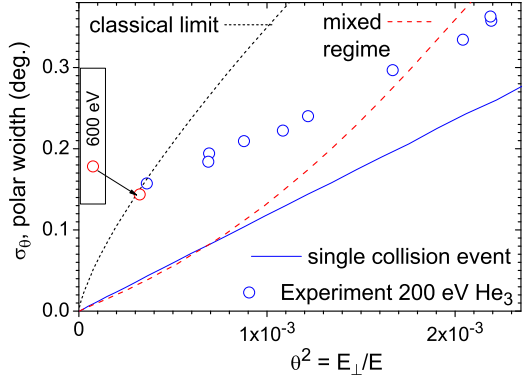}
	\caption{(color online) The polar widths measured in Fig.\ref{fig_y05_Py_fit} are plotted as a function of the ratio $E_\perp /E$ and compared with the prediction of the classical limit (Eq.\ref{sigma_cl}), of the quasi elastic limit  (Eq.\ref{sigma_sc}) and of the mixed inelastic regime (Eq.\ref{sigma_ine_moy}). \label{fig_sig_polar}}
\end{figure}

According to the present model, a significant elastic diffraction probability indicates a quasi-elastic regime where only few collisions are inelastic. The width should then follow Eq.\ref{sigma_sc} and scale linearly with $E_\perp/E$.   
The rms widths of the inelastic profiles are displayed in Fig.\ref{fig_sig_polar}. Once again, the comparison with prediction is far from quantitative. The most salient disagreement being that the experiment widths indicate a minimum value of 0.13 deg. This could be due to the limited surface quality, either microscopic in the form of a reduced mean terrace dimension or macroscopic, in the form of mosaic domain\cite{Lalmi} which was indeed present on some part of the crystal but difficult to identify due to the limited resolution. In this context, the prediction of Eq.\ref{sigma_cl},\ref{sigma_ine_moy} and \ref{sigma_sc} are only plotted to illustrate the distinct angular dependences associated with these three simple regimes.

\subsection{Transverse inelastic line profiles}

\begin{figure}	\includegraphics[width=0.85\linewidth]{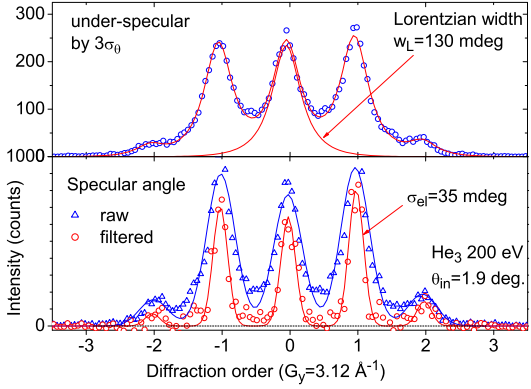}
	\caption{Transverse momentum distribution on the Laue circle (bottom). The intensity is plotted with and without application doubly differential filter suppressing the inelastic contribution. The full lines are fit by gaussian profiles indicating the elastic standard deviation $\sigma_\theta$ =35 mdeg. Data from ref.\cite{Rousseau_2007}.	\label{fig_y05b21Px}}
\end{figure}

\begin{figure}	\includegraphics[width=0.85\linewidth]{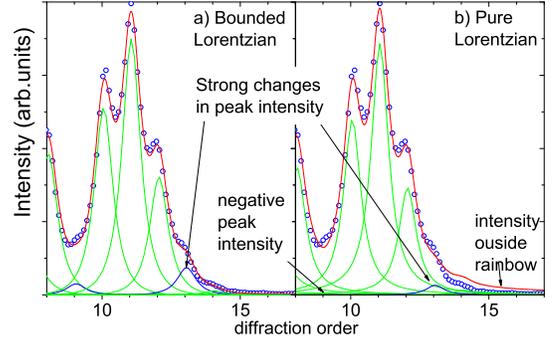}
	\caption{For a diffraction pattern recorded in condition where no elastic intensity is found the inelastic profiles close to the rainbow angle are analyzed by a Bounded Lorentzian profile in a) (see text) and by pure Lorentzian profile in b).\label{peakprofiles}}
\end{figure}

All published analysis of the diffracted intensity have focused on the Laue circle but since the inelastic contribution can not be neglected the question of the peak profile in general and of the inelastic contribution in particular are not well defined. Some empirical descriptions\cite{Rousseau_2008,Seifert_2015} have been proposed which do not take into account the intensity away from the Laue circle and can not compare with the present description.
As shown on the 2D plots of Fig.\ref{circles} the elastic profile on the Laue circle can be isolated by a doubly differential filter. 
The 1D profiles corresponding to Fig.\ref{fig_Images_diff}b) is plotted in the lower part of Fig.\ref{fig_y05b21Px} and displays a strong gaussian character with a width $\sigma_\theta$ corresponding to that of the primary beam. 
This however does not give an answer to the inelastic line profile because the filter is not considered fully quantitative. 
Assuming that the elastic contribution drops rapidly away from the Laue circle, the transverse profiles are analyzed in the upper part of the Fig.\ref{fig_y05b21Px} at a distance of 3 standard deviations $\sigma_\theta$ from the Laue circle. 
This later is well fitted by a Lorentzian profile as used in ref.\cite{Seifert_2015} but with little precision whether the wings are correct or not. The Fig.\ref{fig_Lorentz} suggests that the Lorentzian curve possesses too large wings, this specific aspect is evaluated in Fig.\ref{peakprofiles} recorded along the $<110>$ direction. 
Fig.\ref{peakprofiles}b) shows indeed that the large wings of Lorentzian profile produce significant intensity above the rainbow angle and negative intensities when a diffraction order with low intensity is located in between more intense peaks. Since the profile calculated in the model and displayed in Fig.\ref{fig_Lorentz} are not analytic we have used a simple but empirical "bounded Lorentzian" profile which resembles a standard Lorentzian $L_w(x)=A/(x^2+w^2/4)$ in its center but with wings attenuated by a Gaussian function $BL_w(x)=L_w(x)\times e^{-x^2/4w^2}$. In this case the variance is well defined $\sigma_{BL}\sim 0.732.w$ contrary to the Cauchy-Lorentz distribution.

To further investigate the disagreement pointed above, that inelastic scattering width does not tend to zero in the quasi-elastic regime, we have analyzed the data of Ref.\cite{DebiossacPRL} recorded with 460 eV He$_4$ atoms at 1 deg. with an angular resolution of 7 mdeg and where no trace of surface mosaicity was found (Fig.\ref{fig_Images_diff}a)).

\begin{figure}	\includegraphics[width=0.85\linewidth]{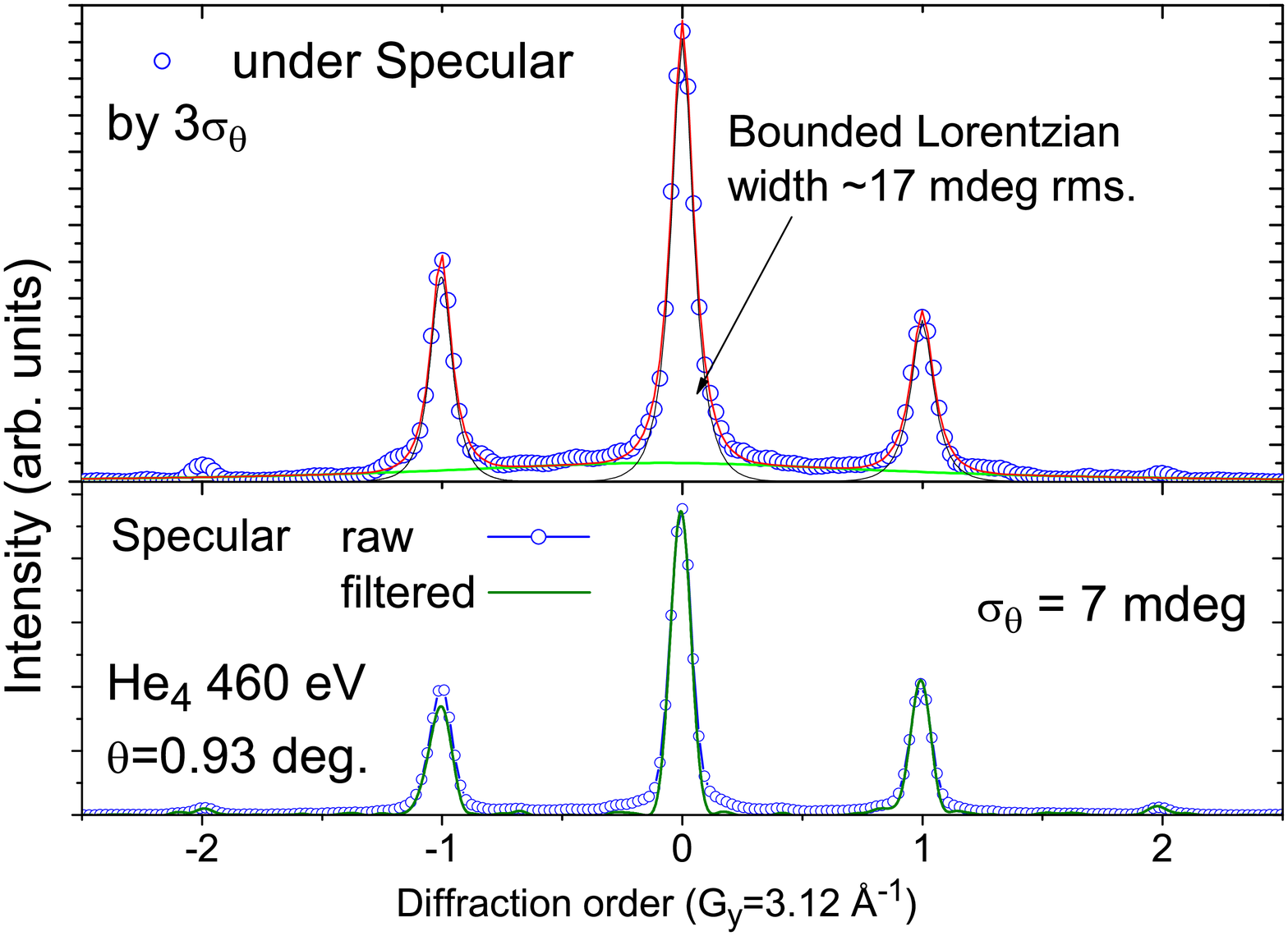}
	\caption{Transverse momentum distribution ($k_y$) at the specular angle (bottom) showing a gaussian profile with $\sigma = 7$ mdeg identical to the primary beam\cite{DebiossacPRL}. The top curve is recorded $24$ mdeg below and should be inelastic. The sharp components have a "bounded Lorentzian" width of 17 mdeg.	\label{fig_y13Px}}
\end{figure}

The elastic ratio is still limited to 50\% but the width of polar profile is lower around 80 mdeg (not shown) which is still more than predicted in Eq.\ref{sigma_sc}. 
Fig.\ref{fig_y13Px} shows the corresponding transverse momentum profiles with narrow elastic peaks on the Laue circle even without application of the doubly differential filter. The inelastic profile is made of sharp peaks on top of a comparatively broad base. It is hard to tell what is the best line profile but the bounded Lorentzian profile (see above) indicate a rms width of 17 mdeg. 
It is tempting to identify this narrow structure as the inelastic profile originating from the quasi-elastic collisions predicted here while the broader contributions would originate from surface defects. At 1 degree incidence a single terrace edge prevents elastic diffraction over a distance $L\gtrsim a/\theta\approx 200\AA$ so that the useful fraction of a terrace of size $T$ is only $(T-L)/T$. All the projectiles getting closer to the terrace edge undergo more and more violent collisions. Since a terrace edge perpendicular to the beam does not affect the transverse periodicity a whole range of inelastic diffraction conditions can be produced.

It should be mentioned that a quasi linear increase of the transverse width, qualitatively in-line with Eq.\ref{sigma_sc} and Eq.\ref{sigma_ine_moy} was measured by \cite{Seifert_2015}. However this observation is performed at the specular angle with an unknown composition of elastic and inelastic contribution and the absence of a clear definition of the reported width prevents a direct comparison. Interestingly though, they suggest that the transverse width is identical along different directions of the crystal surface.

As to the ratio close to three predicted in Fig.\ref{fig_deflection_function} for the for the $k_z$ to $k_y$ inelastic broadening, it can be observed on Fig.\ref{fig_Images_diff}c) and \ref{fig_Images_diff}d) as the elongation ratio of the elliptic diffraction spots. In these figures, the elastic contribution is negligible and  the brighter spots correspond to maximum of the diffraction curve ($dI_m/dk_{eff} \sim 0$) so that the polar profile is hardly distorted by the slowly varying modulation of the Bessel function $J_m$ along this $<100>$ direction. Also, at least in the quasi-elastic regime, the strict correlation between individual scattering angle and energy loss can be considered exact so that the quasi linear increase of the transverse width with $E_\perp$ is also present as $\delta k_y\propto k_{eff}^2$ in each inelastic diffraction image. In other words the spots are slightly distorted ellipses.

\section{Discussion}

We review here some of the assumptions made in the model developed above.

At the very heart of the multiple inelastic model is the local value of the stiffness $\Gamma$ of the repulsive binary interaction potential.
In itself the existence of such binary potentials is not a severe requirement, these can be regarded as an efficient way to fit the 3D potential energy landscape usually evaluated within density functional theory. 

Similarly the requirement that these binary interaction potentials can be expanded as a leading exponential term $V(r)=V e^{-\Gamma r}$ for distances to the surface $z=z_t\pm\sigma_z$ around the turning point $z_t$ is probably not a severe constraint at limited surface temperature. 
Most of the deflection takes place at these distances where the projectile spends most of the interaction time $\tau=(\Gamma v_\perp)^{-1}$ and where the resulting forces are almost perpendicular to the trajectory. 
The other fundamental input of the model is the quantum motion of the surface atoms represented by the surface Debye temperature $T_D$. It enters both the inelastic scattering properties and the Debye-Waller factor. In other words the model presented tries to answer the question: how much momentum can the vibrational wave-function absorb without changing energy?

It turned out that even for the well investigated He-LiF system, the literature reports slightly different values of $\Gamma$\cite{Winter_2002,Boato_1976,Celli_85,Winter_PSS_2011} and surprisingly scattered values of $T_D$ \cite{Pfandzelter,Schuller_PRA2010,Vidali_88} fitted to reproduce specific classical or quantum features.
A critical re-analysis of the Debye surface temperature is needed and would help improving the present model by putting clear boundaries to the quantitative agreement.
The surface Debye temperature is of particular importance for the contribution of the Li atoms completely neglected here. On the rigid lattice the Li atoms are hidden behind the significantly larger F atoms but their light mass could helps them spilling out even at room temperature if $T_D$ is low enough.

The VdW attraction also has been neglected and is certainly important in the 10-50 meV\cite{DebiossacPRB_2014,Diaz_2016,Gravielle_2013,Miraglia_2017} in the form of the Beeby correction\cite{Beeby_1971} where the effective impact energy is $E_\perp=E\sin^2\theta + E_{VdW}$ with $E_{VdW}$ the depth of the potential energy well. The VdW attraction also influences the shape of the binary potential $V(r)$ and may affect the local value of $\Gamma(E_\perp)$ as illustrated in \cite{Gravielle_2014,Miraglia_2017}.

As pointed out in section II, all theoretical descriptions of elastic diffraction start from the potential energy landscape. In opposition to Ref\cite{Aigner_2008,Schuller_PRA2010}, the present model suggests that this PEL should be evaluated with surface atoms at their equilibrium positions without taking into account the thermal motion in the averaging. Starting from this PEL and the projectile mass, the elastic diffraction intensities $I_m(k_{in})$ are calculated by standard method \cite{Rousseau_2007,Zugarramurdi_2012,Manson2008,Diaz_2016b,Gravielle_2013,Sanz} without any surface dynamical property. 
In the present model the effect of the thermal motion is evaluated at each collision to predict a self consistent contribution to the inelastic scattering profiles and to the DWF adapted to grazing angles in Eq.\ref{Pes}. 

The model uses a rigid lattice so that all trajectories remain identical close to the surface irrespective of the inelastic event taking place. The influence of these events is considered only statistically and in the far field. This is probably the main limitation of the model. For instance, at high surface temperature or at larger perpendicular energies, more violent collisions can occur that could make the associated inelastic trajectory significantly different from the elastic one.

Also the common classical trajectory implicit in the model is not adapted to more specific quantum effect where multiple trajectories are involved, for instance bound state resonances\cite{DebiossacPRL} or quantum reflection\cite{Bum_2010}. 

Only one inelastic process has been considered here, the vibrational excitation of an individual surface atom whereas collective phonon modes are identified in TEAS (see \textit{e.g.}\cite{Kraus,Taleb_2016}). The overall interaction time $\tau\sim 1/\Gamma v_\perp$ with the surface is the same but each collision with a surface atom is $N_{eq}$ (Eq.\ref{Neq}) times shorter. The validity of this sudden approximation and the boundary between these regime remains to be investigated. Note that even at hyperthermal energies recent measurements and analysis\cite{Taleb_2017} have suggested that a multiphonon excitation regime can be present with consequences having similarities with the model developed here.

\section{Summary and Conclusion}

A simple and parameter free model has been presented. It describes the multiple smooth collisions taking place along the classical trajectory with surface atoms located at their equilibrium position (rigid lattice). Considering the short collision time, a sudden approximation is developed where only the local Debye oscillator is considered instead of the phonon branches. 
The successive virtual binary recoil energies cumulated along the trajectory (Eq.\ref{Eloss}) is proposed as a criterion to evaluate the overall elastic scattering probability (Eq.\ref{Pes}). 
Three regimes have been identified: 

(\textit{i}) A quasi-elastic one where almost all collisions take place in the M\"{o}ssbauer-Lamb-Dicke regime and where the observed inelastic properties can be understood as deriving from a single inelastic event. 

(\textit{ii}) A quasi classical regime where almost all collisions are inelastic and where quantum effects only reduce the actual amount of energy loss and angular straggling. 

(\textit{iii}) In between a mixed quantum-classical regime is identified progressively linking the quantum and classical limits.

The model suggests that inelastic diffraction intensity are given by $I_m(k_{eff})$ where $k_{eff}=(|k_{in}|+|k_{out}|)/2$ and $I_m(k_{in})$ are the intensity ratios of elastic diffraction. 
The overall polar angle inelastic scattering profile $P(k_{eff})$ is predicted to follow a log-normal distribution while the transverse momentum profile of each diffraction peak $P(k_y)$ is found to follow a quasi Lorentzian line shape. 
In both directions the widths are governed by the amount of inelastic energy loss drawing a direct link between the well documented nuclear energy loss and the polar angular straggling around the specular angle or Laue circle. On the Laue circle both the elastic and inelastic diffraction contribute resulting in a composite scattering profiles.

For pedagogical purposes a simpler model where the $N_{eq}$ collisions taking place close to the turning point are considered equivalent was presented to illustrate the statistical treatment.

The model naturally merges to the classical scattering regime where the cubic dependence $E_{loss} \propto\theta_{in}^3$ (Eq.\ref{Eloss}) is in-line with observations, at least for moderate energies angles and surface temperatures\cite{Villette_these,Pfandzelter,Mertens_2000,Winter_2002}. The progressive blurring of the diffraction features is interpreted as a degradation of the visibility when the transverse width $\sigma_y$ exceeds the Bragg peak separation $G_y$. 

Quantitatively, the comparison with existing experimental data is not fully convincing. The measured elastic fraction is much less than predicted, particularly at the lowest angles of incidence where other sources of decoherence such as topological defects are suspected to contribute.

This underlines an important difficulty inherent to inelastic diffraction that different decoherence mechanisms tend to have comparable consequences often preventing unambiguous interpretation.
Note also that the comparison with scattering profiles is much more demanding than a 2D color plot where a general impression of good agreement is easier to reach.

Several analytical formula have been derived allowing simple estimates of the effect of the change in primary beam energy or angle of incidence, the projectile mass or target mass as well as the sensitivity to temperature and thermal motion.

From the experimental point of view, procedures have been suggested to analyze inelastic diffraction images taking the primary beam as a reference for diffraction circles. The data suggest that the topological defects becoming increasingly important at grazing incidence are the most important limitation to investigate the fully elastic regime.

The $<100>$ direction investigated here is simple in term of modeling but the random direction investigated by Ref\cite{Seifert_2015} produces a simpler scattering profiles and it would be interesting to adapt the present model to random direction.


More work is needed with new data and new binary interaction potentials to be able to discuss the validity of the model and the underlying assumptions.

Finally, several other inelastic processes can contribute to the inelastic signal.
High energy ($\sim$ 10 eV) localized electronic excitations were found to give rise to a momentum exchange larger than the reciprocal lattice vector destroying diffraction\cite{Lienemann_2011,Winter_PSS_2011}. At variance, more gentle electron-hole pair excitation at the Fermi edge of metal\cite{Bunermann_2015} seem to preserve the diffraction features\cite{Bundaleski_2008,Khemliche_2008,Bundaleski_2011} \footnote{An error occurred Fig.3 of\cite{Bundaleski_2011} when reproducing the Fig.7 of\cite{Rousseau_2008} where an inversion of the fit to the data makes an artificially perfect agreement, the error was then reproduced in Fig.3.31 of ref.\cite{Winter_PSS_2011}.}.
Although not discussed here, it is worth mentioning that electronic excitation should also play a role in the elastic scattered intensity. It has been investigated theoretically for molecular projectiles in Ref.\cite{Diaz_2009}.  
So far, fast atomic diffraction on molecular layer\cite{Seifert_alanine} have shown only inelastic behavior and this remain to be investigated more closely and the present model offers a direct link to triangulation approaches\cite{Nataliya,Feiten_2015}.

\section{Acknowledgment}
Several of the ideas presented here were developed during the PHD theses of J. Villette, the internship of K. Cordier, the visit of J.R. Manson and N. Bundaleski. We are also grateful to the successive collaborators starting with P. Rousseau who observed the first diffraction with fast atoms in 2004, H. Khemliche who promoted the patent\cite{Khemliche_patent} as well as A. Momeni,  P. Soulisse and B. Lalmi  with a special attention to the wonderful team operating the MBE  at Institut des Nanosciences de Paris P. Atkinson, M. Eddrief and V. Etgens. Last but not least, continuous en-lightening discussions with A.G. Borisov, J.P. Gauyacq and C. Lesech are kindly acknowledged. A.J. Mayne and P. Atkinson are warmly thanked for their cristicisms and corrections to this manuscript.


\end{document}